\documentclass{aa}  
\usepackage{graphicx}
\usepackage{txfonts}
\begin{document} 
   \title{MCDHF and RCI calculations of energy levels, lifetimes, and transition rates in  \mbox{\ion{Si}{iii} and \ion{Si}{iv}}  }
   \author{B. Atalay
          \inst{1,2}
          \and
          T. Brage\inst{1,4}\
          \and
            P. J\"{o}nsson\inst{3}\          
          \and
          H. Hartman\inst{3}\
          }
   \institute{ Department of Physics, Lund University, Post Office Box 118, SE-22100 Lund, Sweden\\
             \email{betul.atalay@teorfys.lu.se, batalay@comu.edu.tr}
              \and
             Department of Physics, \c{C}anakkale Onsekiz Mart University,
              \c{C}anakkale, Turkey
         \and
             Materials Science and Applied Mathematics, Malm\"{o} University, SE-20506 Malm\"{o}, Sweden
         \and
             Institute of Modern Physics, Fudan University, Shanghai, PRC\\
             \           
             }
   \date{Received-, 2018; }
%
% \abstract{}{}{}{}{} 
% 5 {} token are mandatory
% 
  \abstract
   {We present extensive multiconfiguration Dirac-Hartree-Fock and relativistic configuration interaction calculations including 106 states in doubly ionized silicon (\ion{Si}{III}) and 45 states in triply ionized silicon (\ion{Si}{IV}), which are important for astrophysical determination of plasma properties in different objects. These calculations represents an important extension and improvement of earlier calculations especially for \ion{Si}{III}. The calculations are in good agreement with available experiments for excitation energies, transition properties, and lifetimes. Important deviations from the NIST-database for a selection of perturbed Rydberg series are discussed in detail.}


   \maketitle
%
%________________________________________________________________

\section{Introduction}

  Silicon is one of the most abundant elements on Earth and in the Universe. Due to its high abundance, the lowest ionization ions of Silicon, \ion{Si}{i-iv}, play important roles in the diagnostics and modeling of various astrophysical \citep{Becker90,CATANZARO08, Iijima08} and laboratory plasmas \citep{COWPE08, YAMAZAKI09, OGILVIE09}.
  \ion{Si}{i} and \ion{Si}{ii} have been extensively studied both experimentally and theoretically, see for example \cite{Pehlivan2018} and references given therein.

Lines from \ion{Si}{iii} and \ion{Si}{iv} are observed in the spectra of several different astronomical objects, for example the solar corona and transition region, early-type stars, planetary nebulae, novae, and the interstellar medium, where they have been used for diagnostics of the plasma parameters \citep{Dufton83, Nussbaumer86, Rubin93}.
As an example, \citet{Dufton83} used the \ion{Si}{iii} emission lines observed by Skylab to determine the electron density of the solar corona. Discrepancies between observed and predicted \ion{Si}{iii} line ratios involving the intercombination line $ 3s^2$ $^1S_0$ - $3s3p$  $^3P_1$ at 1892 {\AA}  and the higher excitation $3s3p$ $^{1}P_1$ - $3s4s$ $^1S_0$ at 1313 {\AA} were found. The authors suggested it may have been caused by the presence of a non-Maxwellian electron distributions changing the intensity ratio between the lines. The energy of the upper level of the 1313 {\AA} line is considerably higher than the average electron thermal energy, and the excitation from the ground state can be increased by high-energy electrons (\cite{Pinfield99}. Similarly, \cite{Keenan89a} presented Spacelab 2 observations and \cite{Pinfield99} presented spectra from the Solar Ultraviolet Measurements of Emitted Radiation (SUMER) instrument on-board the Solar and Heliospheric Observatory (SOHO) to give a deeper analysis of possible non-Maxwellian emission-line enhancement in different solar regions. The suggestion of the presence of non-Maxwellian electron distributions was followed up by \citet{Dzif11}. More recently, \citet{DelZanna2015} investigated the main spectral diagnostics for \ion{Si}{iii} ultraviolet lines, to measure electron densities and temperatures. However, they found no conclusive evidence for the presence of non-Maxwellian electron distributions from observations of the low transition region of the solar atmosphere, based on $R$-matrix scattering calculations for electron collisional excitation of \ion{Si}{iii}, carried out with the intermediate-coupling frame transformation method. The updated and more accurate atomic data of \ion{Si}{iii} in the present paper, can be used for rederivation of the plasma diagnostics using existing solar observations.

Due to their relatively high ionization energies, lines from \ion{Si}{iii} and \ion{Si}{iv} appear in early-type stars, for example B-type stars, which are massive stars with surface temperatures 10\,000–30\,000 K showing strong \ion{Si}{iii} and \ion{Si}{iv} lines in their optical spectra. The present-day chemical abundances of the Galaxy in the solar vicinity, can be determined by studying B-type main sequence stars due to their short lifetimes \citep{Przybilla08, Simon10}. The silicon ionization balance is used as the temperature diagnostics in the B-type star temperature range. \citet{Becker90}  used \ion{Si}{iii} and \ion{Si}{iv} spectral lines to determine the temperature, whereas \citet{Monteverde00} studied the ionization equilibria of \ion{Si}{iii} compared to \ion{Si}{ii} and \ion{Si}{iv}, respectively, to find the Si-abundance. 
A recent study by \cite{NievaPrzybilla2012} analyzed 29 early B-type stars in OB associations
with a thorough and self-consistent analysis technique. They found the present-day Solar neighborhood to be chemically homogeneous, indicating abundance fluctuations of less than 10 \% around the mean. \citet{Bailey13} carried out a study on the determination of the abundance of Si to clarify discordant results in mid to late B-type stars. 

The atomic data parameters, particularly transition rates, are essential in the abundance determination and plasma modeling. The work presented here therefore aims to provide accurate atomic data for a large part of the \ion{Si}{iii} and \ion{Si}{iv} spectrum.

\section{Previous work on atomic data for \ion{Si}{iii} and \ion{Si}{iv}} 

\citet{Berry71} and \citet{Livingston:76, Livingston76} performed radiative lifetime measurements in \ion{Si}{iii} and \ion{Si}{iv} using the beam-foil technique. \citet{Kwong83} used a radio-frequency ion trap to measure the lifetime of the long-lived $3s3p$ $^{3}P^{o}_{1}$ level of \ion{Si}{iii} by observation of the decay to the ground state at 1892 {\AA} and obtained an $A$-value of $(1.67\pm 0.1)\times 10^4$~s$^{-1}$. This result has been included in many later plasma models.  Later studies, for instance \citet{Ojha88}, confirm this value.

On the theoretical side, \citet{Nussbaumer86} used a multiconfiguration approach with an adjustable Thomas-Fermi potential to determine the transition probabilities for the 17 lowest terms in \ion{Si}{iii}. They applied various semi-empirical corrections based on the observed energies to improve the accuracy in the term energies. 
\citet{Butler93} computed the radiative data for the $3l3l^\prime$ transitions, for \ion{Si}{iii} and other Mg-like ions of astrophysical interests by using the close-coupling approximation with a modified version of the $R$-Matrix code as a part of the international Opacity project \citep{Seaton87}. \citet{Safronova00} used relativistic second-order many body perturbation theory (MBPT) to calculate excitation energies and transition rates for the same transition arrays. \citet{Almaraz00} applied the CIV3 code \citep{HIBBERT1975} to calculate energy levels and oscillator strengths in \ion{Si}{iii}.

In more recent studies, \citet{DelZanna2015} performed a large-scale $R$-matrix scattering calculation providing 149 $LS$ terms and 283 fine-structure levels arising from 3s$nl$, 3p$nl$ and 3d$nl$ configurations with $n \leq 5$ and $l \leq 4$. 
\cite{AGGARWAL17} reported energies and lifetimes for the 141 levels of the $3l3l^\prime$ and $3l4l^\prime$
configurations and radiative rates for four types of transitions (E1, E2, M1, and M2) in  \ion{Si}{iii}. \citet{IORGA18} investigated the effect of core-valence and core-core correlations on the energy levels and transition probabilities in the Mg isoelectronic sequence using the Flexible Atomic Code (FAC) \citep{Gu2008}.
The results contained the energy of the levels arising from the valence configurations along with transition rates corresponding to E1, M1, E2, M2 transitions between states arising from $3l3l^\prime$ with $l,l^\prime \leq 2 $ and $3snl^{\prime\prime}$  with $n\leq7$ and $l^{\prime\prime} \leq 4$ configurations.

\citet{FroeseFischer06} performed extensive and highly accurate multiconfiguration Hartree-Fock (MCHF) calculations for the Mg-like sequences ($Z=12,...,26$). They used Breit-Pauli approximation to include relativistic effects. They computed both allowed (E1) and some forbidden (E2, M1, M2) transitions.

For triply ionized silicon \citet{MANIAK1993} performed radiative lifetime measurements of $3p, 3d,$ and $4s$ levels using the beam-foil technique.
\citet{Theodosiou88} used a semi-empirical quantum defect approach to calculate lifetimes for the $3p$ $^{2}P_{1/2}$, $3p$ $^{2}P_{3/2}$, $3d$ $^{2}D_{3/2}$, and $3d$ $^{2}D_{5/2}$ levels in the Na isoelectronic sequence. These calculations reached a fair agreement with experiments, but deviated from some ab initio results. \citet{SIEGEL1998} used a semi-empirical model potential approach to compute electric dipole transition oscillator strengths for low-lying transitions in the Na isoelectronic sequence (\ion{Na}{i} - \ion{Ca}{x}). The core polarization effects were explicitly included in the calculations. \citet{SIEMS2001} presented oscillator strengths and lifetimes for \ion{Si}{iv} using a multiconfiguration Hartree-Fock relativistic (HFR) approach. 
In recent years, \citet{Nandy2015} carried out calculations of the relativistic sensitivity coefficients, oscillator strengths, transition probabilities, lifetimes, and magnetic dipole hyperfine structure constants for a number of low-lying states in the \ion{Zn}{ii}, \ion{Si}{iv}, and \ion{Ti}{iv}. \citet{Safronova98} carried out all-order relativistic many-body calculations of removal energies and hyperfine constants for $3s, 3p_{1/2}, 3p_{3/2}, 3d_{3/2},3d_{5/2}$, and $4s$ states of Na-like ions with $Z = 11-16$. The reduced dipole matrix elements were determined for $3p_{1/2}-3s$, $3p_{3/2}-3s$, $4s-3p_{1/2}$, $4s-3p_{3/2}$, $3d_{3/2}-3p_{1/2}$, $3d_{3/2}-3p_{3/2}$, and $3d_{5/2}-3p_{3/2}$ electric-dipole transitions. The calculations included single and double excitations of the Hartree-Fock ground state to all orders in perturbation theory. Theoretical fine-structure intervals had an agreement with
measurements to about 0.3\% for $3p$ states and to about 3\% for $3d$ states. Theoretical hyperfine constants and line strengths agreed with the precise measurements to better than 0.3\%. 
Finally, \citet{FroeseFischer06} performed extensive ab initio non-orthogonal spline CI calculations for the Na-like sequence. \citet{Kelleher08} presented the latest compilation of \ion{Si}{iii} and \ion{Si}{iv} transition rates.

\section{Theory }
In this work, the calculations were performed using the fully relativistic multi-configuration Dirac-Hartree-Fock (MCDHF) method in \(jj\)-coupling \citep{grant2007relativistic,Fischer2016}.

\subsection{Multiconfiguration Dirac-Hartree-Fock}
An electronic state of a many electrons system is determined by a wave function \(\Psi\), which is a solution to the wave equation:
\begin{equation}
H\Psi=E\Psi,
\end{equation}
where \(H\) is the Hamiltonian operator and \(E\) is the total energy of the system. The common starting point of the MCDHF method is the Dirac-Coulomb Hamiltonian:
\begin{equation}
%\label{test}
H_{DC}= \sum_{i=1}^N  \Big (c\boldsymbol{\alpha}_i\cdot\boldsymbol{p_i}+( \beta_i -1)c^2+V_{nuc}(r_i)\Big)+\sum_{i>j}^N \frac{1}{r_{ij}},
\end{equation}
where \(V_{nuc}(r_i)\) is the the monopole part of the electron-nucleus Coulomb interaction, \(r_{ij}\) is the distance between electrons \emph{i} and \emph{j},  \(\boldsymbol{\alpha}\) and \(\beta\) are the 4-by-4 Dirac matrices, and \(c\) is the speed of light.

The approximate solutions to the wave equations  are referred to as atomic state functions (ASFs). An atomic state function, \(\Psi(\gamma PJ)\), is in our approach represented by a linear combination of configuration state functions(CSFs),
\begin{equation}
\Psi(\gamma PJ)=\sum_{j=1}^{NCSFs}c_j\Phi(\gamma_j PJ),
\end{equation}
where \(P\) is the parity and \(J\) is the total angular momentum. \(\gamma_j\) represents all necessary quantum numbers and the orbital occupancy to define the CSF, while \(c_j\) are the mixing coefficients.
 The $\gamma$ is usually selected as the $\gamma_j$ corresponding to the largest weight $|c_j|^2$.

The CSFs are in turn constructed as angular-momentum-coupled, anti-symmetrized products of one-electron Dirac-orbitals of the form

\begin{equation}
\psi(r)=\psi_{n\kappa,m}({r})=\frac{1}{r}
\left( 
\begin{array}{cc}
P_{n\kappa}(r)\chi_{\kappa,m}( {\theta,\varphi}) \\ 
\mbox{i}Q_{n\kappa}(r)\chi_{-\kappa,m}( {\theta,\varphi})
\end{array}
\right),
\end{equation}
where $P_{n\kappa}(r)$ and $Q_{n\kappa}(r)$ are the large and small components of the radial wave function, and $\chi_{\pm\kappa,m}(\theta,\varphi)$ are two-component spin-orbit functions.

The extended optimal level (EOL) scheme is used to determine energies and wave functions.
In the EOL scheme, the radial parts of the Dirac orbitals and the expansion coefficients of the targeted states are optimized to
self-consistency by solving the MCDHF equations, which are derived using the variational approach \citep{Dyall1989}. In subsequent relativistic configuration interaction (RCI) calculations the transverse photon interaction (Breit interaction) 

\begin{equation}
\begin{aligned}
H_{Breit}=&-\sum_{i<j}^N \Bigg[\boldsymbol\alpha_i\cdot \boldsymbol\alpha_j\frac{\cos(w_{ij}r_{ij}/c)}{r_{ij}} \\ &+ \left(\boldsymbol\alpha_i\cdot \boldsymbol\nabla_i\right)\left(\boldsymbol\alpha_i\cdot \boldsymbol\nabla_j\right)
\frac{\cos(w_{ij}r_{ij}/c)-1}{w_{ij}^2r_{ij}/c^2}\Bigg]
\end{aligned}
\end{equation}
may be included in the Hamiltonian \citep{McKenzie1980}. The photon frequency $w_{ij}$, used by the RCI program in calculating the matrix elements of the transverse photon interaction, is taken as the difference of the diagonal Lagrange multipliers associated with the orbitals. 
The leading quantum electrodynamic (QED) corrections effects, in the form of self-energy and vacuum polarization, are also included.
The CSFs are given in the $jj$-coupling scheme during this procedure, but to make a comparison with experiments more feasible, we transform the resulting wave function to the 
\(LSJ\)-coupling scheme, using the JJ2LSJ program \citep{Gaigalas03,Gaigalas04,Gaigalas_17} part of the GRASP2K code \citep{Jonsson13}. All the calculations were performed with an
updated parallel version of the GRASP2K code by \citet{Jonsson13}. To calculate the spin-angular part of the matrix elements, the second quantization method in coupled tensorial form and quasispin technique \citep{Gaigalas1997,Gaigalas2001} were adopted. 

\subsection{Computation of transition parameters}
Once well-converged and effectively complete ASFs have been obtained radiative transition such as transition probabilities and weighted oscillator strengths can be determined. The transition parameters between two states \(\gamma\)\(JM\) and \(\gamma^{\prime}\)\(J^{\prime}M^{\prime}\) are expressed in terms reduced matrix elements
\begin{equation}
\big \langle \Psi(\gamma PJ)||\mathbf{T}||\Psi(\gamma^\prime P^\prime J^\prime)\big \rangle =
\sum_{j,k}c_jc_k^\prime \big \langle \Phi(\gamma_j PJ)||\mathbf{T}||\Phi(\gamma_k^\prime P^\prime J^\prime) \big \rangle ,
\end{equation}
where \(\mathbf{T}\) is the transition operator \citep{Grant74}.

For electric multipole transitions, there are two forms of the transition operator, the length (Babushkin gauge) and velocity (Coulomb gauge) forms \citep{Grant74}. Due to the definition of these two, the length form is more sensitive to the outer part of the wave functions, which are usually active in radiative transitions.  A number of studies have shown that the length form generally gives more reliable values at a given level of valence and core–valence electron correlation although the velocity form seems to be more stable for transitions including highly excited Rydberg states \citep{Pehlivan17}.
It is common to use the agreement between transition rates $A_{l}$ and $A_{v}$ computed in two forms as an indicator of accuracy of the wave functions, (see review by \citet{FroeseFischer2009, Ekman14}). A possible measure of this is the quantity \(dT\), characterizing the uncertainty of the calculated transition rates and defined as
\begin{equation}
%\label{test}
dT= \frac{\left|A_{l}-A_{v}\right|}{\text{max}({A_{l},A_{v}})},
\end{equation}
where $A_l$ and $A_v$ are the transition rates in length and velocity forms respectively. The values of $dT$ do not represent uncertainty estimates for each individual transition but should be considered as statistical indicators of uncertainties within given sets of transitions.

\section{Calculations}

Our MCDHF and RCI calculations for doubly and triply ionized silicon started by defining a multireference (MR) set of configurations. From this we allow single and double substitutions to a systematically increasing active set of orbitals.

\subsection{\ion{Si}{iii}}
In doubly ionized silicon, calculations were performed for states belonging to the
$3s^2$, the $3p^2$; the $3p4p$; the $3sns$ with $n=4,...,9$; the $3snd$ with $n=3,...,8$; the $3sng$ with $n=5,...,8$ even configurations and furthermore, the $3snp$ with $n=3,...,9$; the $3snf$ with $n=4,...,8$; the $3pnd$ with $n=3,4$; and the $3p4s$ odd configurations. These configurations define the MR for the even and odd parities, respectively. Terms involving configurations with $n = 8$ and $9$ do not belong to our targeted states but they are taken into account to obtain orbitals with large radii to get a reasonable agreement between length and velocity form for the transition properties, see \citet{Pehlivan17}.

In the first step of our calculations, an initial MCDHF calculation in the EOL scheme \citep{Dyall1989} was performed simultaneously for all even and odd multireference states. These initial calculations were followed by calculations with expansions including the configuration state functions (CSFs) obtained by single (S), double (D) substitutions of electrons from the spectroscopic reference configurations to the active set of orbitals \citep{Olsen1988, Sturesson2007}.

In an restricted active set approach (RAS) restrictions are put on the allowed substitutions from the MR, when generating the full space of CSFs. 
We therefore define the valence region of the atom as the two outer electrons, outside the $1s^22s^22p^6$ core subshells. We kept the $1s^2$-subshell fixed in all calculations, that is, not allowing substitutions from it. 
After optimizing simultaneously the even and odd states of the MR-set in the first step of calculations, our goal was to include valence-valence (VV) and core-valence (CV) interaction to convergence. We optimized four layers of correlation orbitals based on the VV correlation, only allowing SD substitutions from the valence subshells. The orbitals in the active set were systematically extended to include orbitals up to the $13s,13p,12d,12f,12g$, and $9h$ in the final correlation layer. These MCDHF calculations were followed by RCI calculations including Breit-interaction and some QED effects as described above. 

As a final step of our work, an RCI calculation was performed. The expansion for that RCI calculation was obtained by augmenting the largest SD valence expansion with a CV expansion. The CV expansion was generated by SD substitutions from the valence orbitals and the $2p^6$ core with the restriction that there should be at the most one substitution from the $2s$ or $2p$ subshells. We neglected core-core (CC) correlation, meaning more than one excitation from the core, which was comparatively unimportant for both the energy separations and the transition probabilities \citep{Zou_Fischer00}. 
The resulting expansions consisted of 1\,401\,150 and 1\,760\,209 CSFs distributed over the $J= 0,1,...,6$ symmetries for even and odd parity, respectively.

\subsubsection{\ion{Si}{iv}}
In triply ionized silicon, calculations were performed for states belonging to the configurations $2s^22p^6nl$ where $n\leq 9$ and $l\leq 6$, defining the MR configurations. 
We again added two more layers of spectroscopic orbitals, $n=8,9$, in comparison with the states we were targeting to get a reasonable agreement between length and velocity form for the transition properties.

As a starting point for the calculations, MCDHF calculations in the EOL scheme were performed for the even and odd parity states in the MR simultaneously. The initial calculations were followed by separate calculations in the EOL scheme for the even and odd parity states, where the CSFs were obtained by allowing SD substitutions from the configurations in the MR to active orbital sets, which were consecutively enlarged by layers of correlation orbitals. \ion{Si}{iv} has one electron outside closed subshells, and consequently there is no VV correlation. The corrections to our results can therefore be classified as CV and CC correlations. The major effect comes from the CV correlation.
The CV expansion was obtained by restricting the substitutions in such a way that only one substitution was allowed from the $2s$ or $2p$ subshells of the configurations in the MR, and no substitutions from the $1s$ shell, this means that $1s$ shell was an inactive closed core. 

The active sets of orbitals for the even and odd parity states were extended by layers to include orbitals with quantum numbers $n\leq 12$ and $l\leq 6$. Each MCDHF calculation was followed by RCI calculations, including the Breit-interaction and leading QED effects.
We investigated the CC correlation effects by optimizing one layer of orbitals $n=13$ on SD from the $2s^22p^6$ core in an RCI calculation as a final step of our work.
The number of CSFs in the final even and odd state expansions were approximately 995\,020 and 993\,501, respectively, distributed over the different $J$ symmetries.

\section{Results}
\subsection{\ion{Si}{iii}}
We present in Table \ref{SiIII_energy} the computed excitation energies in \ion{Si}{iii} for increasing active sets of orbitals labeled with the highest principal quantum number $n$ of the orbitals in the active set. The calculations including only VV correlation, built on three layers of correlation orbitals, is followed by finally including the CV correlation using an RCI-approach.

The computed excitation energies are in good agreement with the values from the NIST-database \citep{NIST_ASD}.
For the VV correlation, the mean relative difference between theory and experiment is of the order of 0.9\%. 
The inclusion of CV correlation effects improves the energies dramatically since the values for all computed energy separations have converged.

The final excitation energies are in good agreement with the experimental ones, with a mean difference of only 0.05\%. For comparison, the experimental energies from the NIST Atomic Spectra Database (\cite{NIST_ASD}) and also the differences \(\Delta{E}\), between the observed energies and the final computed excitation energies, are given in Table \ref{SiIII_energy}.

There is excellent agreement between observations and calculations for most of the levels, with a few important exceptions, due to mislabeling of levels. The NIST label classification for \ion{Si}{iii} is based on the analysis done by \citet{Toresson61}. Since then the NIST designations for the $3p3d$ $^{1}P$ and the $3p4s$ $^{1}P$ have been interchanged as recommended by \citet{Victor76} and \citet{Zetterberg_1977}. The assignments for $3p3d$ $^{1}P$ and $3p3d$ $^{1}F$ levels have been questioned previously by \citet{Reistad_1984} and \citet{Tomas_89}.

\begin{figure}[ht]
\centering
\includegraphics[width=\hsize]{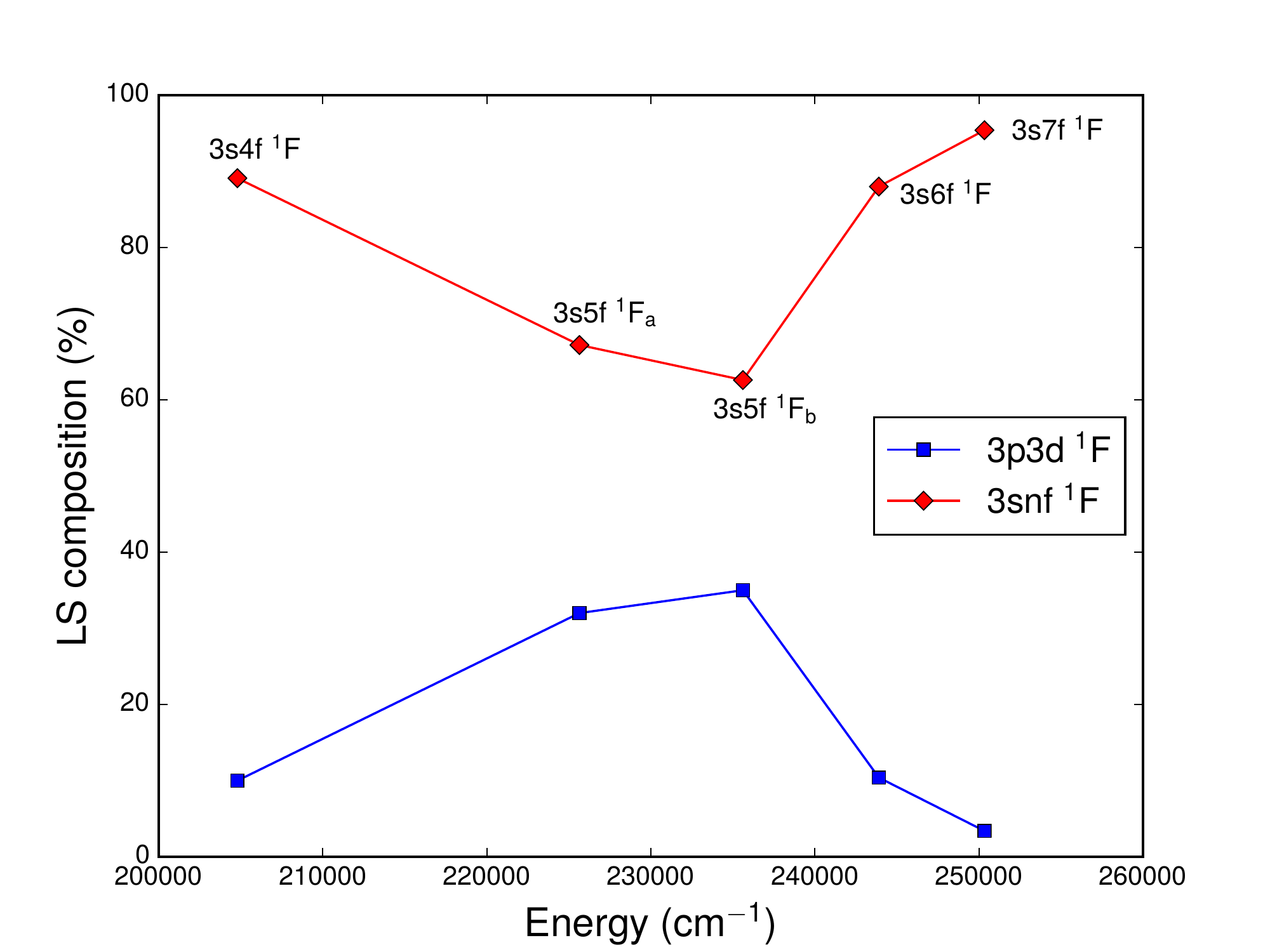}
\caption{ $3p3d$ and $3snf$ composition of $^{1}F^o$ Rydberg series members.}
\label{LS_comp_1F_SiIII}
\end{figure}

If we start by investigating the $^{1}F^{o}$-levels, we note that the $3p3d$ $^{1}F^{o}$  perturbs the $3snf$ Rydberg series. This perturbation has been studied by a number of authors along the Mg-sequence \citep{Tomas_89, Reistad_1984,Aashamar_1986}. The NIST classification identifies the level at 235413 cm$^{-1}$  as $3p3d$ $^{1}F$ for convenience, even though the calculated composition is 65 \% $3snf$ \citep{FroeseFischer82}. To illustrate the complexity of the situation, we show the composition of the $^{1}F$ levels in Figure \ref{LS_comp_1F_SiIII}. It is clear that the $3p3d$ $^{1}F$ is not the major component for any state and that the same CSF $3s5f$ $^{1}F^{o}$ is the largest component for two levels.   We therefore choose to change the designation as shown in Table \ref{SiIII_energy} where these two levels are labeled as $3s5f ^{1}F^{o}_{3 ~a}$ and $3s5f ^{1}F^{o}_{3 ~b}$, respectively.
%%%%%%%%%%%%%%%%%%%%%%%%%%%%%%%%%%%%%%%%%%%%%%%%%%%%%%%%%%%%%%%%%%%%%%%%%%%%%%%%%%%%%
%%%%%%%%%%%%%%%%%%%%%%%%%%%%%%%%%%%%%%%%%%%%%%%%%%%%%%%%%%%%%%%%%%%%%%%%%%%%%%%%%%%%%
 
\begin{figure}[ht]
\centering
\includegraphics[width=\hsize]{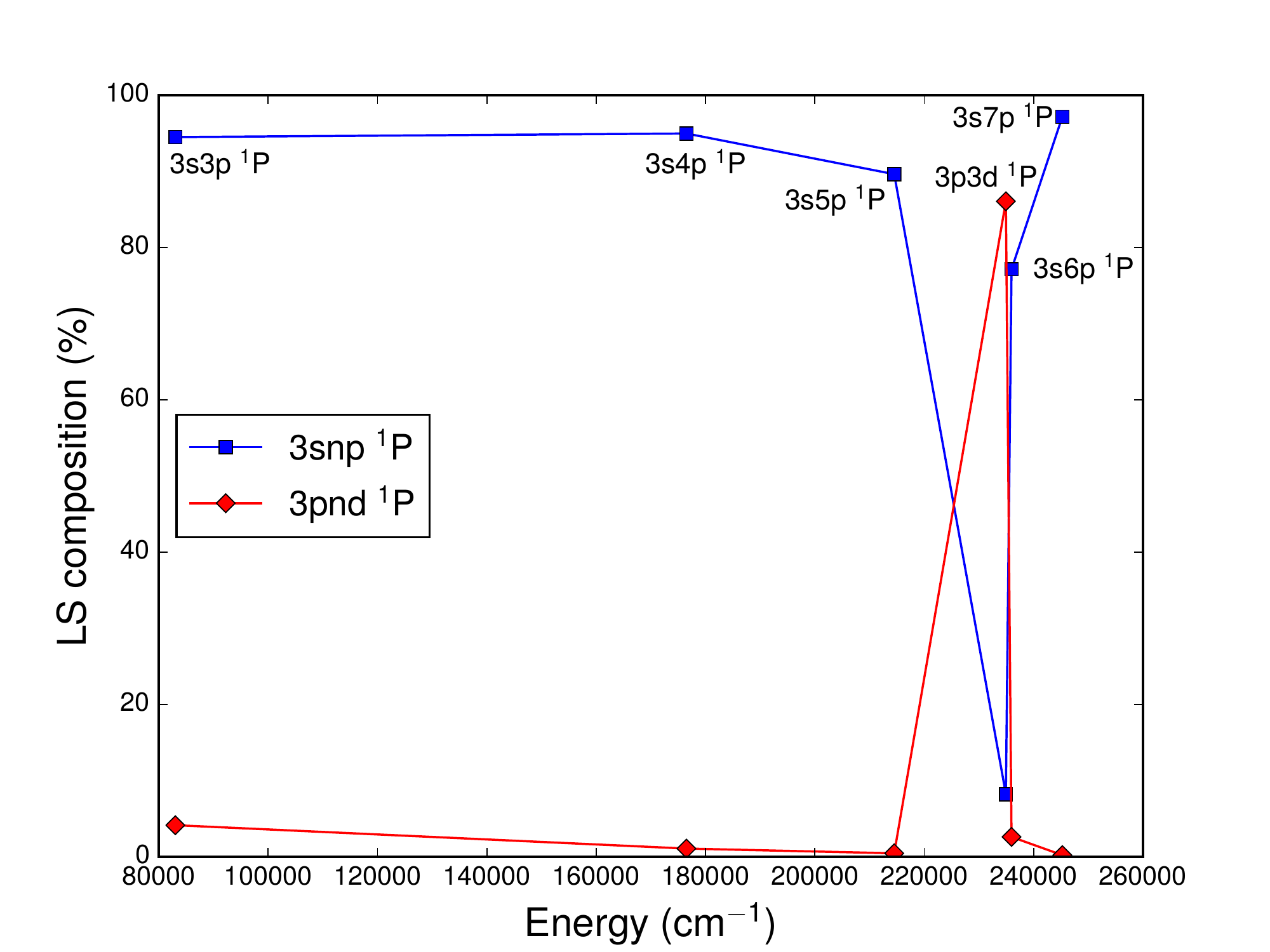}
\caption{$3pnd$ $^{1}P$ and $3snp$ composition of $^{1}P^o$ Rydberg series members.}
\label{LS_comp_1P_SiIII}
\end{figure}

 Our second example of a strongly perturbed series investigated by \citet{FroeseFischer82} is a short-range interaction of the plunging configuration ${3p3d}$ ${^{1}P}$ with the ${3snp}$ $^{1}P$ Rydberg series. In \ion{Si}{iii} the $3s6p$ $^{1}P_1$ and $3p3d$ $^{1}P_1$ are close to degenerate and the labeling of these levels is hard to reproduce. A closer look at the $LS$-percentage composition and excitation energies  of $3snp$ $^{1}P$ and $3p3d$ $^{1}P$ are shown in Figure \ref{LS_comp_1P_SiIII}, where the level at 234923 cm$^{-1}$ is best represented as $3p3d$ $^{1}P$. Also in this case we adjust the labels from NIST. 
 
 A third example of a complicated case of level-designation is the $3snd$ $^{1}D^e$ Rydberg series, which is perturbed by the $3p^2$ $^1D$ for low excitations and the $3p4p$ $^{1}D_2$ for $n=6-8$. In our calculations the level energy of $3s7d$ $^{1}D_2$ is 247946 cm$^{-1}$, while the values given by NIST is 250636 cm$^{-1}$. However, the $LS$ composition of the former level is 72\% $3s7d$ $^{1}D_2$ and 18\% $3p4p$ $^{1}D_2$. We suggest that the NIST identification of the level $^{1}D_2$ at 250636 cm$^{-1}$ and  at 247935 cm$^{-1}$ should be interchanged as shown in Table \ref{SiIII_energy}.

%%%%%%%%%%%%%%%%%%%%%%%%%%%%%%%%%%%%%%%%%%%%%%%%%%%%%%%%%%%%%%%%%%%%%%%%%%%%%%%%%%%%%
%%%%%%%%%%%%%%%%%%%%%%%%%%%%%%%%%%%%%%%%%%%%%%%%%%%%%%%%%%%%%%%%%%%%%%%%%%%%%%%%%%%%%
A problem in spectroscopic calculations, that includes a large number of states, is to reproduce very close degeneracy between some levels. An example of this is the relative position of the singlet and triplet levels the $3snp$ Rydberg series with $n=6-7$. It is clear that we are not able to reproduce the NIST-values for the relative position of these levels, which makes, for example, our values for intercombination lines inaccurate from the $3s6p$ and $3s7p$ $^{3}P$-levels.

In Table \ref{SiIII_energy_comp}, the current results for the 29 lowest excitation energies of \ion{Si}{iii} are compared with the ones from the MCHF-BP calculations by \cite{FroeseFischer06} and $R$-matrix calculations by \citet{DelZanna2015}. To make the comparison easier the differences between observed and computed energies are also given in the last columns for the different computational approaches. From the comparison, it is clear that the agreement with the present RCI calculations and the MCHF-BP calculations is very good. Furthermore, the agreement between the current RCI results and the observed energies is improved especially above the $3p^2$ $^{1}S_{0}$ level.

The complete transition data table, using the length form of transition operator, including rates, weighted oscillator strengths, wavenumber, and the uncertainty in the computed rates, for E1 transitions in \ion{Si}{iii} for the wavelength range of 40-18239 nm is available on-line. The wavenumber and wavelength values are changed to match the values in the NIST database, which are the values from \citet{Reader80}. Uncertainties of the transition rates have been estimated from the expressions suggested by \citet{Ekman14}. The estimated relative uncertainties for most of the strong transitions are below 1\%. For many weak transitions the uncertainties are around 10\%. However, for weak intercombination transitions the uncertainties are considerably large due to difficulties in calculating the transition rates of intercombination lines. There are some weak two-electron one-photon transitions with large uncertainties. These transitions are zero in the single configuration approximation and are allowed only through correlation effects. Thus, the two-electron one-photon transitions are very difficult to compute and the uncertainties are often large.

The $3s4s$ $^{1}S_{0}$ $\rightarrow$ $3s3p$ $^{1}P_{1}$ transition is of special importance in the diagnostics of non-Maxwellian electron  distributions, as discussed above. According to \citet{Dufton83} the rate for this transition is $2.8 \times 10^8$ s$^{-1}$, while \citet{Nussbaumer86} predicted $4.0 \times 10^8$ s$^{-1}$ and \citet{DelZanna2015} $2.96 \times 10^8$ s$^{-1}$.
Our result for this rate is $8.53 \times 10^8$ s$^{-1}$ close to \citet{FroeseFischer06} calculation of $8.99 \times 10^8$ s$^{-1}$.

Our estimated lifetimes (including only E1-transitions) for excited states of \ion{Si}{iii} are given in Table \ref{SiIII_lifetime_comp}. We also compare our results to available measurements in this table. \citet{Berry71}, \citet{Livingston76} and \citet{Bashkin80} measured the lifetimes for a few levels in \ion{Si}{iii} with the beam foil method. The agreement between the current RCI results and measurements is quite satisfactory, though the results from the measurements of  \citet{Berry71} slightly differ from the RCI values. This may be explained by the uncertainties in the beam-foil method discussed by \citet{Zou1999}.
The lifetimes from the current RCI calculations are also compared to results from the MCHF-BP calculations by Froese Fischer et al. (2006) and MBPT calculations by \cite{Safronova00}, when these include all important transitions from a given state. The overall agreement is good, giving us confidence in the present results. For most states there is no major discrepancy between the current results and MBPT results except for the  $3p3d$ $^{3}F_{2,3,4}^{0}$ states. The reason for that is unclear, but for the rest of the states the discrepancies are quite small.

\subsection{\ion{Si}{iv}}

In Table \ref{SiIV_energy}, we present the level energies of the lowest 45 levels in \ion{Si}{iv} as functions of increasing active sets of orbitals (labeled by the highest principal quantum number $n$). The calculations are based on CV correlation, including three layers of correlation orbitals. From the inspection of Table \ref{SiIV_energy}, it is clear that the present calculations are well converged with respect to the increasing orbital set. The inclusion of CC correlation effects, adding one more layer of correlation orbitals with $n=13$, improves the energies and the final energies are in good agreement with experiment displaying a mean difference of 0.09\%. We present the experimental level energies from the NIST-database for \ion{Si}{iv} in Table \ref{SiIV_energy}, including the differences between the observed and the computed energies. In Table \ref{SiIV_energy}, we also compare the present computed excitation energies with those from the non-orthogonal B-spline CI method \citep{Zatsarinny_2002} by \citet{FroeseFischer06} for \ion{Si}{iv}, showing excellent agreement.

The complete transition data table, using the length form of transition operator, including rates, weighted oscillator strengths, wavenumber, and the uncertainty in the computed rates, for E1 transitions in \ion{Si}{iv} for the wavelength range of 31-11933 nm is available on-line. Uncertainties of the transition rates have been estimated from the expressions suggested by \citet{Ekman14}. The wavelength and wavenumber values are changed to match the values in the NIST database, which are the values from \citet{Reader80}. In general, the estimated relative uncertainties for most of the transitions are less than 10\%. However, for some transitions the uncertainties are very large mostly due to difficulties in calculating the transition rates of intercombination lines and two-electron transitions.

 Table \ref{SiIV_lifetime_comp} presents our estimated lifetimes of the excited states (including only E1-transitions) in both length and velocity forms. The average relative difference between lifetimes in two forms is $\sim$ 0.8\%, which is highly satisfactory. This difference might be considered to be an internal validation of the accuracy of the calculations. The comparison of our results with previously reported theoretical and experimental values are also given in Table \ref{SiIV_lifetime_comp}. \citet{SIEMS2001} performed a HFR approach to estimate the lifetimes for \ion{Si}{iv}. In that approach, the electrostatic parameters were optimized using a least-squares procedure. \citet{Berry71} and later \citet{Bashkin80} measured lifetimes using the beam-foil method. The overall agreement between these measured lifetimes and the current lifetimes is rather good. From the table, one can see that our results are in excellent agreement with the results of \citet{FroeseFischer06} and \citet{SIEMS2001}.

%
%______________________________________________________________

\section{Conclusions}
In this work, we performed self-consistent MCDHF and subsequent RCI calculations for \ion{Si}{iii} and \ion{Si}{iv}, as an extension and update to earlier calculations. Previous theoretical and experimental data were used to validate our results. Excitation energies from the RCI calculations are in a very good agreement with available observations. Our present study is an extension to the most accurate MCHF-BP and the BSR\textunderscore CI calculations, and in general in excellent agreements with them. The presented results significantly increase the amount of accurate energy data of astrophysical interest for the two Si-ions. The highly accurate atomic data helps to correct interpretation of the lines. Therefore, we recommend our present results based on a fully relativistic method for abundance analysis and plasma diagnostics.

  % \begin{enumerate}
    %  \item T.
      
   %\end{enumerate}
\begin{acknowledgements}
   Bet{\"u}l Atalay acknowledges financial support from the Scientific and Technological Research Council of Turkey (TUBITAK) - BIDEB 2219 International Post-Doctoral Research Fellowship Program. Tomas Brage, Per J\"{o}nsson, and Henrik Hartman acknowledge support from the Swedish Research Council (VR) under contract No.2015-04842. Dr Atalay also would like to express her appreciation of the hospitality shown by the Division of Mathematical Physics at Lund University and by the Department of Materials Science and Applied Mathematics at Malm\"{o} University.
    
\end{acknowledgements}

%-------------------------------------------------------------------

%%%%%%%%%%%%%%%%%%%%%%%%%%%%%%%%%%%%%%%%%%%%%%%%%%%%%%%%%%%%%%%%%%%%%%%%%%%%%%%%%%%%%%%%%%%%%%%%%%%%%%%%%%%%%%%%%%%%%%%%%%%

%%%%%%%%%%%% 	TABLES  	%%%%%%%%%%%%%%%%%%%%%%%%%%%%%%%%%%%%

%%%%%%%%%%%%%%%%%%%%%%%%%%%%%%%%%%%%%%%%%%%%%%%%%%%%%%%%%%%%

%%%%%%%%%%%%%%%%%%%%%%%%%%%%%%%%%%%%%%%%%%%%%%%%%%%%%%%%%%%%

%%%%%%%%%%%% 	Energy Tables  Si III 	%%%%%%%%%%%%%%%%%%%%%%%%%%%%%%%%%%%%

%%%%%%%%%%%%%%%%%%%%%%%%%%%%%%%%%%%%%%%%%%%%%%%%%%%%%%%%%%%%

%%%%%%%%%%%%%%%%%%%%%%%%%%%%%%%%%%%%%%%%%%%%%%%%%%%%%%%%%%%%%%%%%%%%%%%%%%%%%%%%%%%%%%%%%%%%%%%%%%%%%

%%%%%%%%%%%%%%%%%%%%%%%%%%%%%%%%%%%%%%%%%%%%%%%%%%%%%%%%%%%%

%%%%%%%%%%%% Transition Rates Tables  Si III  Online	%%%%%%%%%%%%%%%%%%%%%%%%%%%%%%%%%%%%

%%%%%%%%%%%%%%%%%%%%%%%%%%%%%%%%%%%%%%%%%%%%%%%%%%%%%%%%%%%%
\onllongtab{

\label{Transitiondata_SiIII}
\begin{longtable}{l l l l l c c c c} 
\caption{Presentation of the complete computed radiative transition data of \ion{Si}{iii} lines together with the relative uncertainties $dT=\frac{\left|A_{l}-A_{v}\right|}{\text{max}({A_{l},A_{v}})}$. }\\ 
\hline\hline 
Upper & Lower & {$\sigma$ (cm$^{-1}$)} &{$\lambda^{vacuum}$ (nm)} & $A$ (s$^{-1}$) 	&	 $gf$ 	& $A_{resc}$ (s$^{-1}$)	&  $gf_{resc}$  & $dT$                      \\ 
\hline
\endfirsthead
\caption{continued.}\\
\hline\hline
Upper & Lower & {$\sigma$ (cm$^{-1}$)} &{$\lambda^{vacuum}$ (nm)} & $A$ (s$^{-1}$) 	&	 $gf$ 	& $A_{resc}$ (s$^{-1}$)	&  $gf_{resc}$  & $dT$                                           \\ 
\hline
\endhead
\hline
\endfoot

$3s7p~^1P_{  1  }^o$ 	&	 $3s^2 ~^1S_{  0  }$        	&	244871.47	&	40.8378	$^*$	&	5.920E+07	&	4.426E-03	&	5.893E+07	&	4.419E-03	&	0.057	\\
$3s7p~^3P_{  1  }^o$ 	&	 $3s^2 ~^1S_{  0  }$        	&	244933.36	&	40.8274	$^*$	&	2.200E+04	&	1.650E-06	&	2.202E+04	&	1.650E-06	&	0.009	\\
$3s6p~^1P_{  1  }^o$ 	&	 $3s^2 ~^1S_{  0  }$        	&	234387.64	&	42.6644	$^*$	&	3.250E+07	&	2.625E-03	&	3.186E+07	&	2.608E-03	&	0.058	\\
$3p3d~^1P_{  1  }^o$ 	&	 $3s^2 ~^1S_{  0  }$        	&	235951.09	&	42.3817	$^*$	&	1.555E+07	&	1.268E-03	&	1.576E+07	&	1.274E-03	&	0.058	\\
$3s6p~^3P_{  1  }^o$ 	&	 $3s^2 ~^1S_{  0  }$        	&	234428.4	&	42.6569	$^*$	&	1.719E+04	&	1.408E-06	&	1.720E+04	&	1.408E-06	&	0.108	\\
$3p4s~^1P_{  1  }^o$ 	&	 $3s^2 ~^1S_{  0  }$        	&	228699.75	&	43.7255	$^*$	&	5.023E+07	&	4.316E-03	&	5.017E+07	&	4.314E-03	&	0.001	\\
$3p4s~^3P_{  1  }^o$ 	&	 $3s^2 ~^1S_{  0  }$        	&	226527.42	&	44.1448	$^*$	&	1.532E+05	&	1.342E-05	&	1.531E+05	&	1.342E-05	&	0.044	\\
$3s5p~^3P_{  1  }^o$ 	&	 $3s^2 ~^1S_{  0  }$        	&	214994.65	&	46.5128	$^*$	&	7.148E+04	&	6.954E-06	&	7.147E+04	&	6.954E-06	&	0.038	\\
\hline
\end{longtable}
\tablefoot{Wavelength and wavenumber values are from the NIST database \citep{NIST_ASD} when available.  Wavelengths marked with $^*$ are the Ritz wavelengths calculated from the energy differences and these transitions are not identified in the NIST database. }\\
}
% End onllongtab

%%%%%%%%%%%%%%%%%%%%%%%%%%%%%%%%%%%%%%%%%%%%%%%%%%%%%%%%%%%%

%%%%%%%%%%%% Transition Rates Tables  Si IV  Online	%%%%%%%%%%%%%%%%%%%%%%%%%%%%%%%%%%%%

%%%%%%%%%%%%%%%%%%%%%%%%%%%%%%%%%%%%%%%%%%%%%%%%%%%%%%%%%%%%
\onllongtab{

\label{Transitiondata_SiIV}
\begin{longtable}{l l l l l c c c c} 
\caption{Presentation of the complete computed radiative transition data of \ion{Si}{iv} lines together with the relative uncertainties $dT=\frac{\left|A_{l}-A_{v}\right|}{\text{max}({A_{l},A_{v}})} $. } \\
\hline \hline 
Upper & Lower & {$\sigma$ (cm$^{-1}$)} &{$\lambda^{vacuum}$ (nm)} & $A$ (s$^{-1}$) 	&	 $gf$ 	& $A_{resc}$ (s$^{-1}$)	&  $gf_{resc}$  & $dT$                      \\ 
\hline
\endfirsthead
\caption{continued.}\\
\hline\hline
Upper & Lower & {$\sigma$ (cm$^{-1}$)} &{$\lambda^{vacuum}$ (nm)} & $A$ (s$^{-1}$) 	&	 $gf$ 	& $A_{resc}$ (s$^{-1}$)	&  $gf_{resc}$  & $dT$                                           \\ 
\hline
\endhead
\hline
\endfoot
$7p~^2P_{ 3/2  }^o$ 	&	 $3s~^2S_{ 1/2  }$   	&	322338	&	31.0234		&	1.046E+08	&	6.047E-03	&	1.049E+08	&	6.052E-03	&	0.071	\\
$7p~^2P_{ 1/2  }^o$ 	&	 $3s~^2S_{ 1/2  }$   	&	322313	&	31.0257		&	1.070E+08	&	3.093E-03	&	1.073E+08	&	3.096E-03	&	0.069	\\
$6p~^2P_{ 3/2  }^o$ 	&	 $3s~^2S_{ 1/2  }$   	&	305682	&	32.7137		&	1.733E+08	&	1.114E-02	&	1.738E+08	&	1.115E-02	&	0.019	\\
$6p~^2P_{ 1/2  }^o$ 	&	 $3s~^2S_{ 1/2  }$   	&	305641	&	32.7181		&	1.773E+08	&	5.702E-03	&	1.778E+08	&	5.707E-03	&	0.017	\\
$5p~^2P_{ 3/2  }^o$ 	&	 $3s~^2S_{ 1/2  }$   	&	276579	&	36.156		&	2.801E+08	&	2.200E-02	&	2.809E+08	&	2.202E-02	&	0.004	\\
$5p~^2P_{ 1/2  }^o$ 	&	 $3s~^2S_{ 1/2  }$   	&	276504	&	36.1659		&	2.872E+08	&	1.129E-02	&	2.880E+08	&	1.130E-02	&	0.003	\\
$7d~^2D_{ 3/2  }$   	&	 $3p~^2P_{ 1/2  }^o$ 	&	256074	&	39.0511		&	7.351E+07	&	6.733E-03	&	7.369E+07	&	6.738E-03	&	0.092	\\
$7d~^2D_{ 5/2  }$   	&	 $3p~^2P_{ 3/2  }^o$ 	&	255613	&	39.1216		&	9.078E+07	&	1.252E-02	&	9.099E+07	&	1.253E-02	&	0.083	\\
\hline
\end{longtable}
\tablefoot{Wavelength and wavenumber values are from the NIST database \citep{NIST_ASD} when available.  Wavelengths marked with $^*$ are the Ritz wavelengths calculated from the energy differences and these transitions are not identified in the NIST database. }\\
}
% End onllongtab

%%%%%%%%%%%%%%%%%%%%%%%%%%%%%%%%%%%%%%%%%%%%%%%%%%%%%%%%%%%%%%%%%%%%%%%%%%%%%%%%%%%%%%%%%%%%%%%%%%%%%%%%%%%%%%%%%%%%%%%%%%%%%%%%%%%%%%%%%%%
\bibliography{bibli.bib}

\begin{thebibliography}{69}
\expandafter\ifx\csname natexlab\endcsname\relax\def\natexlab#1{#1}\fi

\bibitem[{Aashamar {et~al.}(1986)Aashamar, Luke, \& Talman}]{Aashamar_1986}
Aashamar, K., Luke, T.~M., \& Talman, J.~D. 1986, Physica Scripta, 34, 386

\bibitem[{Aggarwal(2017)}]{AGGARWAL17}
Aggarwal, K.~M. 2017, Atomic Data and Nuclear Data Tables, 117-118, 320

\bibitem[{Almaraz {et~al.}(2000)Almaraz, Hibbert, Lavín, Martín, \&
  Bell}]{Almaraz00}
Almaraz, M.~A., Hibbert, A., Lavín, C., Martín, I., \& Bell, K.~L. 2000,
  Journal of Physics B: Atomic, Molecular and Optical Physics, 33, 3277

\bibitem[{{Bailey} \& {Landstreet}(2013)}]{Bailey13}
{Bailey}, J.~D. \& {Landstreet}, J.~D. 2013, \aap, 551, A30

\bibitem[{{Bashkin} {et~al.}(1980){Bashkin}, {Astner}, {Mannervik},
  {Ramanujam}, {Scofield}, {Huldt}, \& {Martinson}}]{Bashkin80}
{Bashkin}, S., {Astner}, G., {Mannervik}, S., {et~al.} 1980, \physscr, 21, 820

\bibitem[{{Becker} \& {Butler}(1990)}]{Becker90}
{Becker}, S.~R. \& {Butler}, K. 1990, \aap, 235, 326

\bibitem[{Berry {et~al.}(1971)Berry, Bromander, Curtis, \& Buchta}]{Berry71}
Berry, H.~G., Bromander, J., Curtis, L.~J., \& Buchta, R. 1971, Physica
  Scripta, 3, 125

\bibitem[{Brage \& Hibbert(1989)}]{Tomas_89}
Brage, T. \& Hibbert, A. 1989, Journal of Physics B: Atomic, Molecular and
  Optical Physics, 22, 713

\bibitem[{Butler {et~al.}(1993)Butler, Mendoza, \& Zeippen}]{Butler93}
Butler, K., Mendoza, C., \& Zeippen, C.~J. 1993, Journal of Physics B: Atomic,
  Molecular and Optical Physics, 26, 4409

\bibitem[{Catanzaro {et~al.}(2008)Catanzaro, Leone, Busá, \&
  Romano}]{CATANZARO08}
Catanzaro, G., Leone, F., Busá, I., \& Romano, P. 2008, New Astronomy, 13, 113

\bibitem[{Cowpe {et~al.}(2008)Cowpe, Astin, Pilkington, \& Hill}]{COWPE08}
Cowpe, J., Astin, J., Pilkington, R., \& Hill, A. 2008, Spectrochimica Acta
  Part B: Atomic Spectroscopy, 63, 1066 , a collection of papers presented at
  the Euro Mediterranean Symposium on Laser Induced Breakdown Spectroscopy
  (EMSLIBS 2007)

\bibitem[{{Del Zanna} {et~al.}(2015){Del Zanna}, {Fern{\'a}ndez-Menchero}, \&
  {Badnell}}]{DelZanna2015}
{Del Zanna}, G., {Fern{\'a}ndez-Menchero}, L., \& {Badnell}, N.~R. 2015, \aap,
  574, A99

\bibitem[{{Dufton} {et~al.}(1983){Dufton}, {Hibbert}, {Kingston}, \&
  {Doschek}}]{Dufton83}
{Dufton}, P.~L., {Hibbert}, A., {Kingston}, A.~E., \& {Doschek}, G.~A. 1983,
  \apj, 274, 420

\bibitem[{{Dyall} {et~al.}(1989){Dyall}, {Grant}, {Johnson}, {Parpia}, \&
  {Plummer}}]{Dyall1989}
{Dyall}, K.~G., {Grant}, I.~P., {Johnson}, C.~T., {Parpia}, F.~A., \&
  {Plummer}, E.~P. 1989, Computer Physics Communications, 55, 425

\bibitem[{{Dzif{\v{c}}{\'a}kov{\'a}} \& {Kulinov{\'a}}(2011)}]{Dzif11}
{Dzif{\v{c}}{\'a}kov{\'a}}, E. \& {Kulinov{\'a}}, A. 2011, \aap, 531, A122

\bibitem[{{Ekman} {et~al.}(2014){Ekman}, {Godefroid}, \& {Hartman}}]{Ekman14}
{Ekman}, J., {Godefroid}, M., \& {Hartman}, H. 2014, Atoms, 2, 215

\bibitem[{Fischer \& Godefroid(1982)}]{FroeseFischer82}
Fischer, C.~F. \& Godefroid, M. 1982, Physica Scripta, 25, 394

\bibitem[{{Froese Fischer}(2009)}]{FroeseFischer2009}
{Froese Fischer}, C. 2009, Physica Scripta Volume T, 134, 014019

\bibitem[{{Froese Fischer} {et~al.}(2016){Froese Fischer}, {Godefroid},
  {Brage}, {J{\"o}nsson}, \& {Gaigalas}}]{Fischer2016}
{Froese Fischer}, C., {Godefroid}, M., {Brage}, T., {J{\"o}nsson}, P., \&
  {Gaigalas}, G. 2016, Journal of Physics B Atomic Molecular Physics, 49,
  182004

\bibitem[{{Froese Fischer} {et~al.}(2006){Froese Fischer}, {Tachiev}, \&
  {Irimia}}]{FroeseFischer06}
{Froese Fischer}, C., {Tachiev}, G., \& {Irimia}, A. 2006, Atomic Data and
  Nuclear Data Tables, 92, 607

\bibitem[{{Gaigalas} {et~al.}(2017){Gaigalas}, {Fischer}, {Rynkun}, \&
  {J{\"o}nsson}}]{Gaigalas_17}
{Gaigalas}, G., {Fischer}, C., {Rynkun}, P., \& {J{\"o}nsson}, P. 2017, Atoms,
  5, 6

\bibitem[{{Gaigalas} {et~al.}(2001){Gaigalas}, {Fritzsche}, \&
  {Grant}}]{Gaigalas2001}
{Gaigalas}, G., {Fritzsche}, S., \& {Grant}, I.~P. 2001, Computer Physics
  Communications, 139, 263

\bibitem[{{Gaigalas} {et~al.}(1997){Gaigalas}, {Rudzikas}, \& {Froese
  Fischer}}]{Gaigalas1997}
{Gaigalas}, G., {Rudzikas}, Z., \& {Froese Fischer}, C. 1997, Journal of
  Physics B Atomic Molecular Physics, 30, 3747

\bibitem[{{Gaigalas} {et~al.}(2003){Gaigalas}, {{\v{Z}}alandauskas}, \&
  {Rudzikas}}]{Gaigalas03}
{Gaigalas}, G., {{\v{Z}}alandauskas}, T., \& {Rudzikas}, Z. 2003, Atomic Data
  and Nuclear Data Tables, 84, 99

\bibitem[{{Gaigalas} {et~al.}(2004){Gaigalas}, {Zalandauskas}, \&
  {Fritzsche}}]{Gaigalas04}
{Gaigalas}, G., {Zalandauskas}, T., \& {Fritzsche}, S. 2004, Computer Physics
  Communications, 157, 239

\bibitem[{Grant(2007)}]{grant2007relativistic}
Grant, I. 2007, Relativistic Quantum Theory of Atoms and Molecules: Theory and
  Computation, Springer Series on Atomic, Optical, and Plasma Physics
  (Springer)

\bibitem[{{Grant}(1974)}]{Grant74}
{Grant}, I.~P. 1974, Journal of Physics B Atomic Molecular Physics, 7, 1458

\bibitem[{{Gu}(2008)}]{Gu2008}
{Gu}, M.~F. 2008, Canadian Journal of Physics, 86, 675

\bibitem[{Hibbert(1975)}]{HIBBERT1975}
Hibbert, A. 1975, Computer Physics Communications, 9, 141

\bibitem[{{Iijima} \& {Nakanishi}(2008)}]{Iijima08}
{Iijima}, T. \& {Nakanishi}, H. 2008, A\&A, 482, 865

\bibitem[{Iorga \& Stancalie(2018)}]{IORGA18}
Iorga, C. \& Stancalie, V. 2018, Atomic Data and Nuclear Data Tables, 123-124,
  313

\bibitem[{{J{\"o}nsson} {et~al.}(2013){J{\"o}nsson}, {Gaigalas}, {Biero{\'n}},
  {Fischer}, \& {Grant}}]{Jonsson13}
{J{\"o}nsson}, P., {Gaigalas}, G., {Biero{\'n}}, J., {Fischer}, C.~F., \&
  {Grant}, I.~P. 2013, Computer Physics Communications, 184, 2197

\bibitem[{{Keenan} {et~al.}(1989){Keenan}, {Cook}, {Dufton}, \&
  {Kingston}}]{Keenan89a}
{Keenan}, F.~P., {Cook}, J.~W., {Dufton}, P.~L., \& {Kingston}, A.~E. 1989,
  \apj, 340, 1135

\bibitem[{{Kelleher} \& {Podobedova}(2008)}]{Kelleher08}
{Kelleher}, D.~E. \& {Podobedova}, L.~I. 2008, Journal of Physical and Chemical
  Reference Data, 37, 1285

\bibitem[{Kramida {et~al.}(2018)Kramida, {Yu.~Ralchenko}, Reader, \& {and NIST
  ASD Team}}]{NIST_ASD}
Kramida, A., {Yu.~Ralchenko}, Reader, J., \& {and NIST ASD Team}. 2018, {NIST
  Atomic Spectra Database (ver. 5.5.6), [Online]. Available:
  {\tt{https://physics.nist.gov/asd}} [2015, April 16]. National Institute of
  Standards and Technology, Gaithersburg, MD.}

\bibitem[{{Kwong} {et~al.}(1983){Kwong}, {Johnson}, {Smith}, \&
  {Parkinson}}]{Kwong83}
{Kwong}, H.~S., {Johnson}, B.~C., {Smith}, P.~L., \& {Parkinson}, W.~H. 1983,
  Physical Review A, 27, 3040

\bibitem[{Livingston {et~al.}(1976{\natexlab{a}})Livingston, Baudinet-Robinet,
  Garnir, \& Dumont}]{Livingston:76}
Livingston, A.~E., Baudinet-Robinet, Y., Garnir, H.~P., \& Dumont, P.~D.
  1976{\natexlab{a}}, J. Opt. Soc. Am., 66, 1393

\bibitem[{Livingston {et~al.}(1976{\natexlab{b}})Livingston, Kernahan, Irwin,
  \& Pinnington}]{Livingston76}
Livingston, A.~E., Kernahan, J.~A., Irwin, D. J.~G., \& Pinnington, E.~H.
  1976{\natexlab{b}}, Journal of Physics B: Atomic and Molecular Physics, 9,
  389

\bibitem[{Maniak {et~al.}(1993)Maniak, Träbert, \& Curtis}]{MANIAK1993}
Maniak, S., Träbert, E., \& Curtis, L. 1993, Physics Letters A, 173, 407

\bibitem[{{McKenzie} {et~al.}(1980){McKenzie}, {Grant}, \&
  {Norrington}}]{McKenzie1980}
{McKenzie}, B.~J., {Grant}, I.~P., \& {Norrington}, P.~H. 1980, Computer
  Physics Communications, 21, 233

\bibitem[{{Monteverde} {et~al.}(2000){Monteverde}, {Herrero}, \&
  {Lennon}}]{Monteverde00}
{Monteverde}, M.~I., {Herrero}, A., \& {Lennon}, D.~J. 2000, \apj, 545, 813

\bibitem[{{Nandy} \& {Sahoo}(2015)}]{Nandy2015}
{Nandy}, D.~K. \& {Sahoo}, B.~K. 2015, \mnras, 447, 3812

\bibitem[{{Nieva} \& {Przybilla}(2012)}]{NievaPrzybilla2012}
{Nieva}, M.~F. \& {Przybilla}, N. 2012, \aap, 539, A143

\bibitem[{{Nussbaumer}(1986)}]{Nussbaumer86}
{Nussbaumer}, H. 1986, \aap, 155, 205

\bibitem[{Ogilvie \& Nicolich(2009)}]{OGILVIE09}
Ogilvie, R.~E. \& Nicolich, J. 2009, Spectrochimica Acta Part B: Atomic
  Spectroscopy, 64, 788 , a Collection of Papers Presented at the 19th
  International Congress on X-Ray Optics and Microanalysis (ICXOM-19)

\bibitem[{{Ojha} {et~al.}(1988){Ojha}, {Keenan}, \& {Hibbert}}]{Ojha88}
{Ojha}, P.~C., {Keenan}, F.~P., \& {Hibbert}, A. 1988, Journal of Physics B
  Atomic Molecular Physics, 21, L395

\bibitem[{{Olsen} {et~al.}(1988){Olsen}, {Roos}, {J{\o}rgensen}, \&
  {Jensen}}]{Olsen1988}
{Olsen}, J., {Roos}, B.~O., {J{\o}rgensen}, P., \& {Jensen}, H.~J.~A. 1988,
  \jcp, 89, 2185

\bibitem[{Pehlivan~Rhodin(2018)}]{Pehlivan2018}
Pehlivan~Rhodin, A. 2018, PhD thesis, Lund University

\bibitem[{{Pehlivan Rhodin} {et~al.}(2017){Pehlivan Rhodin}, {Hartman},
  {Nilsson}, \& {J{\"o}nsson}}]{Pehlivan17}
{Pehlivan Rhodin}, A., {Hartman}, H., {Nilsson}, H., \& {J{\"o}nsson}, P. 2017,
  \aap, 598, A102

\bibitem[{Pinfield {et~al.}(1999)Pinfield, Keenan, Mathioudakis, Phillips,
  Curdt, \& Wilhelm}]{Pinfield99}
Pinfield, D.~J., Keenan, F.~P., Mathioudakis, M., {et~al.} 1999, The
  Astrophysical Journal, 527, 1000

\bibitem[{Przybilla {et~al.}(2008)Przybilla, Nieva, \& Butler}]{Przybilla08}
Przybilla, N., Nieva, M.-F., \& Butler, K. 2008, The Astrophysical Journal
  Letters, 688, L103

\bibitem[{{Reader} {et~al.}(1980){Reader}, {Corliss}, {Wiese}, \&
  {Martin}}]{Reader80}
{Reader}, J., {Corliss}, C.~H., {Wiese}, W.~L., \& {Martin}, G.~A. 1980,
  {Wavelengths and transition probabilities for atoms and atomic ions: Part 1.
  Wavelengths, part 2. Transition probabilities} ({U.S. Government Printing
  Office})

\bibitem[{Reistad {et~al.}(1984)Reistad, Brage, Ekberg, \&
  Engström}]{Reistad_1984}
Reistad, N., Brage, T., Ekberg, J.~O., \& Engström, L. 1984, Physica Scripta,
  30, 249

\bibitem[{{Rubin} {et~al.}(1993){Rubin}, {Dufour}, \& {Walter}}]{Rubin93}
{Rubin}, R.~H., {Dufour}, R.~J., \& {Walter}, D.~K. 1993, \apj, 413, 242

\bibitem[{Safronova {et~al.}(1998)Safronova, Derevianko, \&
  Johnson}]{Safronova98}
Safronova, M.~S., Derevianko, A., \& Johnson, W.~R. 1998, Phys. Rev. A, 58,
  1016

\bibitem[{Safronova {et~al.}(2000)Safronova, Johnson, \& Berry}]{Safronova00}
Safronova, U.~I., Johnson, W.~R., \& Berry, H.~G. 2000, Phys. Rev. A, 61,
  052503

\bibitem[{Seaton(1987)}]{Seaton87}
Seaton, M.~J. 1987, Journal of Physics B: Atomic and Molecular Physics, 20,
  6363

\bibitem[{{Siegel} {et~al.}(1998){Siegel}, {Migdalek}, \& {Kim}}]{SIEGEL1998}
{Siegel}, W., {Migdalek}, J., \& {Kim}, Y.-K. 1998, Atomic Data and Nuclear
  Data Tables, 68, 303

\bibitem[{Siems {et~al.}(2001)Siems, Luna, \& Trigueiros}]{SIEMS2001}
Siems, A., Luna, F., \& Trigueiros, A. 2001, Journal of Quantitative
  Spectroscopy and Radiative Transfer, 68, 635

\bibitem[{{Sim\'on-D\'{\i}az, S.}(2010)}]{Simon10}
{Sim\'on-D\'{\i}az, S.} 2010, A\&A, 510, A22

\bibitem[{{Sturesson} {et~al.}(2007){Sturesson}, {J{\"o}nsson}, \& {Froese
  Fischer}}]{Sturesson2007}
{Sturesson}, L., {J{\"o}nsson}, P., \& {Froese Fischer}, C. 2007, Computer
  Physics Communications, 177, 539

\bibitem[{Theodosiou \& Curtis(1988)}]{Theodosiou88}
Theodosiou, C.~E. \& Curtis, L.~J. 1988, Phys. Rev. A, 38, 4435

\bibitem[{Toresson(1961)}]{Toresson61}
Toresson, Y.~G. 1961, Ark. Fys. (Stockholm), 18, 389

\bibitem[{{Victor} {et~al.}(1976){Victor}, {Stewart}, \& {Laughlin}}]{Victor76}
{Victor}, G.~A., {Stewart}, R.~F., \& {Laughlin}, C. 1976, \apjs, 31, 237

\bibitem[{Yamazaki {et~al.}(2009)Yamazaki, Yoshiki, Takemura, Tomita, \&
  Takeno}]{YAMAZAKI09}
Yamazaki, H., Yoshiki, M., Takemura, M., Tomita, M., \& Takeno, S. 2009,
  Spectrochimica Acta Part B: Atomic Spectroscopy, 64, 808 , a Collection of
  Papers Presented at the 19th International Congress on X-Ray Optics and
  Microanalysis (ICXOM-19)

\bibitem[{Zatsarinny \& Fischer(2002)}]{Zatsarinny_2002}
Zatsarinny, O. \& Fischer, C.~F. 2002, Journal of Physics B: Atomic, Molecular
  and Optical Physics, 35, 4669

\bibitem[{Zetterberg \& Magnusson(1977)}]{Zetterberg_1977}
Zetterberg, P.~O. \& Magnusson, C.~E. 1977, Physica Scripta, 15, 189

\bibitem[{Zou \& Fischer(2000)}]{Zou_Fischer00}
Zou, Y. \& Fischer, C.~F. 2000, Phys. Rev. A, 62, 062505

\bibitem[{{Zou} {et~al.}(1999){Zou}, {Hutton}, {Huldt}, {Martinson}, {Ando},
  {Oyama}, {Nystr{\"o}m}, {Kambara}, \& {Yamazaki}}]{Zou1999}
{Zou}, Y., {Hutton}, R., {Huldt}, S., {et~al.} 1999, Physica Scripta Volume T,
  80, 460

\end{thebibliography}
\bibliographystyle{aa}

\onecolumn
\begin{appendix} 
\section{Additional Tables}
\begin{longtable}{lccccccr} 
\caption{\label{SiIII_energy}Computed excitation energies in cm $^{-1}$ for \ion{Si}{iii} from different computational models. The differences $\Delta E $ between the final computations and the observed values are shown in the last column. }
\\ \hline\hline
&& VV &&&& \\
\cline{2-5}
Level	&	$n=10$	&	$n=11$	&	$n=12$	&	$n=13$	&	CV	&	$E_{\text{{obs}}}$$^a$	&	$\Delta E$	\\
\hline
\endfirsthead
\caption{continued.}\\
\hline\hline
&& VV &&&& \\
\cline{2-5}
Level	&	$n=10$	&	$n=11$	&	$n=12$	&	$n=13$	&	CV	&	$E_{\text{{obs}}}$$^a$	&	$\Delta E$	\\
\hline
\endhead
\hline
\endfoot
$3s^2 ~^1S_{   0   }$        ~  	&	0	&	0	&	0	&	0	&	0	&	0	&	0	\\
$3s3p~^3P_{   0   }^o$ ~  	&	51539	&	51541	&	51561	&	51604	&	52790	&	52725	&	-65	\\
$3s3p~^3P_{   1   }^o$ ~  	&	51656	&	51671	&	51686	&	51728	&	52913	&	52853	&	-60	\\
$3s3p~^3P_{   2   }^o$ ~  	&	51893	&	51938	&	51941	&	51981	&	53164	&	53115	&	-49	\\
$3s3p~^1P_{   1   }^o$ ~  	&	83380	&	83186	&	83145	&	83094	&	83031	&	82884	&	-147	\\
$3p^2 ~^1D_{   2   }$ ~  	&	120291	&	120305	&	120314	&	120361	&	122447	&	122215	&	-232	\\
$3p^2 ~^3P_{   0   }$ ~  	&	128668	&	128658	&	128667	&	128703	&	129832	&	129708	&	-124	\\
$3p^2 ~^3P_{   1   }$ ~  	&	128808	&	128794	&	128796	&	128832	&	129953	&	129842	&	-111	\\
\hline
\end{longtable}
\tablefoot{$^*$ Labeling is changed from NIST-standard to better represent the composition of different levels in a Rydberg series, according to our calculations. As an example are two levels labeled with the same $3s5f~^{1}F$ term but different indices are used to distinguish them. Details are given in the text.}
\tablebib{$^{(a)}$\citet{NIST_ASD}.\\
%$^*$ Labelling is changed from NIST-standard to better represent the composition of different levels in a Rydberg series, according to our calculations. As an example are two levels labeled with the same $3s5f$ $^{1}F^$ term but different indices are used to distinguish them. Details are given in the text.
}
%\end{longtab}

%%%%%%%%%%%% 	Comparison Energy Tables  Si III 	%%%%%%%%%%%%%%%%%%%%%%%%%%%%%%%%%%%%

%%%%%%%%%%%%%%%%%%%%%%%%%%%%%%%%%%%%%%%%%%%%%%%%%%%%%%%%%%%%%%%%%%%%%%%%%%%%%%%%%%%%%%%%%%%%%%%%%

%\begin{longtab}
\begin{longtable}{lccccccc} 
\caption{\label{SiIII_energy_comp} Observed and computed excitation energies for the 29 lowest states in \ion{Si}{iii}, from present calculations ($E_{\text{RCI}}$) and other theoretical results ($E_{\text{theor}}^{c,d}$). These are compared to values from the NIST-database ($E_{\text{obs}}^b$). $\Delta{E}$ represents the difference between observed and computed energies. All energies are given in cm$^{-1}$.}
%\centering
\\ \hline \hline 
Level	&	$E_{\text{RCI}}^a$	&	$E_{\text{obs}}^b$ &	$E_{\text{theor}}^c$	&	$E_{\text{theor}}^d$ &	$\Delta E_{\text{RCI}}^a$	&	$\Delta E_{\text{theor}}^c$ & $\Delta E_{\text{theor}}^d$	\\ \hline
$3s^2 ~^1S_{   0   }$        ~  	&	0	&	0	&	0	&	0	&	0	&	0	&	0	\\
$3s3p~^3P_{   0   }^o$ ~  	&	52790	&	52725	&	52704	&	51156	&	-65	&	20	&	1568	\\
$3s3p~^3P_{   1   }^o$ ~  	&	52913	&	52853	&	52835	&	51267	&	-60	&	18	&	1586	\\
$3s3p~^3P_{   2   }^o$ ~  	&	53164	&	53115	&	53099	&	51491	&	-49	&	16	&	1624	\\
$3s3p~^1P_{   1   }^o$ ~  	&	83031	&	82884	&	83069	&	84060	&	-147	&	-185	&	-1176	\\
$3p^2 ~^1D_{   2   }$ ~  	&	122447	&	122215	&	122487	&	120194	&	-232	&	-273	&	2020	\\
$3p^2 ~^3P_{   0   }$ ~  	&	129832	&	129708	&	129751	&	128551	&	-124	&	-43	&	1157	\\
$3p^2 ~^3P_{   1   }$ ~  	&	129953	&	129842	&	129891	&	128664	&	-111	&	-50	&	1177	\\
$3p^2 ~^3P_{   2   }$ ~  	&	130193	&	130101	&	130153	&	128886	&	-92	&	-53	&	1214	\\
$3s3d~^3D_{   3   }$   ~  	&	143106	&	142944	&	143640	&	142257	&	-162	&	-697	&	686	\\
$3s3d~^3D_{   2   }$   ~  	&	143162	&	142946	&	143638	&	142252	&	-216	&	-693	&	693	\\
$3s3d~^3D_{   1   }$   ~  	&	143208	&	142948	&	143644	&	142249	&	-260	&	-696	&	699	\\
$3s4s~^3S_{   1   }$   ~  	&	153332	&	153377	&	153881	&	151563	&	45	&	-504	&	1814	\\
$3p^2 ~^1S_{   0   }$ ~  	&	153953	&	153444	&	153855	&	155103	&	-509	&	-411	&	-1659	\\
$3s4s~^1S_{   0   }$   ~  	&	159065	&	159070	&	159604	&	159383	&	5	&	-535	&	-314	\\
$3s3d~^1D_{   2   }$   ~  	&	166013	&	165765	&	166490	&	168327	&	-248	&	-725	&	-2562	\\
$3s4p~^3P_{   0   }^o$ ~  	&	175219	&	175230	&	175704	&	174006	&	11	&	-474	&	1224	\\
$3s4p~^3P_{   1   }^o$ ~  	&	175247	&	175263	&	175743	&	174036	&	16	&	-480	&	1227	\\
$3s4p~^3P_{   2   }^o$ ~  	&	175312	&	175336	&	175821	&	174102	&	24	&	-485	&	1234	\\
$3s4p~^1P_{   1   }^o$ ~  	&	176503	&	176487	&	176963	&	175288	&	-16	&	-476	&	1199	\\
$3p3d~^3F_{   2   }^o$ ~  	&	199259	&	198923	&	199701	&		&	-336	&	-778	&		\\
$3p3d~^3F_{   3   }^o$ ~  	&	199318	&	199026	&	199811	&		&	-292	&	-785	&		\\
$3p3d~^3F_{   4   }^o$ ~  	&	199402	&	199164	&	199955	&		&	-238	&	-791	&		\\
$3s4d~^3D_{   1   }$   ~  	&	201691	&	201598	&	202258	&		&	-93	&	-661	&		\\
$3s4d~^3D_{   2   }$   ~  	&	201666	&	201598	&	202263	&		&	-68	&	-665	&		\\
$3s4d~^3D_{   3   }$   ~  	&	201634	&	201599	&	202265	&		&	-35	&	-666	&		\\
$3s4d~^1D_{   2   }$   ~  	&	204464	&	204331	&	205114	&		&	-133	&	-784	&		\\
$3s4f~^1F_{   3   }^o$ ~  	&	204795	&	204829	&	205537	&		&	33	&	-709	&		\\
$3p3d~^1D_{   2   }^o$ ~  	&	205357	&	205029	&	205765	&		&	-328	&	-736	&		\\
\hline \\
\end{longtable}
\tablebib{
$^{(a)}$Present calculations;$^{(b)}$\citet{NIST_ASD};
$^{(c)}$\citet{FroeseFischer06}; $^{(d)}$\citet{DelZanna2015}.
}
%\end{longtab}

%%%%%%%%%%%%%%%%%%%%%%%%%%%%%%%%%%%%%%%%%%%%%%%%%%%%%%%%%%%%

%%%%%%%%%%%% Lifetime Tables  Si III 	%%%%%%%%%%%%%%%%%%%%%%%%%%%%%%%%%%%%

%%%%%%%%%%%%%%%%%%%%%%%%%%%%%%%%%%%%%%%%%%%%%%%%%%%%%%%%%%%%
\clearpage
%\begin{longtab}
\centering
\begin{longtable}{lcccrll} 
\caption{\label{SiIII_lifetime_comp} Results for \ion{Si}{III}: Comparison between computed lifetimes, in length ($\tau_l$) and velocity ($\tau_v$) gauge, from our calculations. These are compared to the predicted lifetimes from MCHF-BP$^b$ ($\tau_{\text{MCHF-BP}}$) and MBPT$^c$  ($\tau_{\text{MBPT}}$) models, as well as experimental results $\tau_{\text{obs}}^{d,e}$, with stated uncertainties. All values are given in seconds. }

\\ \hline \hline
&RCI$^a$&&&&&\\
\cline{2-3}
Level & $\tau_l$ & $\tau_v$ & $\tau_{\text{MCHF-BP}}$$^b$ & $\tau_{\text{MBPT}}$$^c$ & $\tau_{\text{obs}}$$^d$ & $\tau_{\text{obs}}$$^e$\\ \hline
\endfirsthead
\caption{continued.}\\
\hline\hline
&RCI$^a$&&&&&\\
\cline{2-3}
Level & $\tau_l$ & $\tau_v$ & $\tau_{\text{MCHF-BP}}$$^b$ & $\tau_{\text{MBPT}}$$^c$ & $\tau_{\text{obs}}$$^d$ & $\tau_{\text{obs}}$$^e$\\ \hline
\endhead
\hline
\endfoot

$3s3p~^3P_{  1  }^o$ 	&	7.679E-05	&	7.822E-05	&	5.809E-05	&	1.010E-04	&		&		\\
$3s3p~^1P_{  1  }^o$ 	&	4.024E-10	&	4.010E-10	&	4.050E-10	&	4.290E-10	&		& 	$(4 \pm 1)$E-10 	\\
$3p^2 ~^1D_{  2  }$ 	    &	3.438E-08	&	3.331E-08	&	3.273E-08	&	4.570E-08	&	$(2.60 \pm 0.15)$E-08	&	 $(2.6 \pm 0.3)$E-08 	\\
$3p^2 ~^3P_{  0  }$ 	    &	4.747E-10	&	4.726E-10	&	4.779E-10	&	4.650E-10	&	$(5.0 \pm 0.3)$E-10$^f$	&	$(3.4 \pm 1.0)$E-10	\\
$3p^2 ~^3P_{  1  }$ 	    &	4.735E-10	&	4.714E-10	&	4.764E-10	&	4.740E-10	&	$(5.0 \pm 0.3)$E-10$^f$	&	$(3.4 \pm 1.0)$E-10	\\
$3p^2 ~^3P_{  2  }$     	&	4.715E-10	&	4.693E-10	&	4.741E-10	&	4.210E-10	&	$(5.0 \pm 0.3)$E-10$^f$	&	$(3.4 \pm 1.0)$E-10	\\
$3s3d~^3D_{  3  }$   	&	3.564E-10	&	3.564E-10	&	3.524E-10	&	3.770E-10	&		&		\\
$3s3d~^3D_{  2  }$   	&	3.540E-10	&	3.545E-10	&	3.506E-10	&	3.740E-10	&		&		\\
$3s3d~^3D_{  1  }$   	&	3.523E-10	&	3.531E-10	&	3.494E-10	&	3.740E-10	&		&		\\
$3s4s~^3S_{  1  }$   	&	4.069E-10	&	4.078E-10	&	4.068E-10	&		&		&		\\
$3p^2 ~^1S_{  0  }$ 	    &	4.328E-10	&	4.306E-10	&	4.440E-10	&	4.760E-10	&	$(5.8 \pm 0.4)$E-10$^f$	&		\\
$3s4s~^1S_{  0  }$   	&	1.172E-09	&	1.178E-09	&	1.112E-09	&		&		&		\\
$3s3d~^1D_{  2  }$   	&	2.185E-10	&	2.186E-10	&	2.170E-10	&	2.320E-10	&		&		\\
$3s4p~^3P_{  0  }^o$ 	&	3.400E-09	&	3.422E-09	&	3.409E-09	&		&	$(3.3 \pm 0.3)$E-09	&	 $(4.1 \pm 0.5)$E-09	\\
$3s4p~^3P_{  1  }^o$ 	&	3.385E-09	&	3.403E-09	&	3.390E-09	&		&	$(3.6 \pm 0.3)$E-09	&	 $(4.5 \pm 0.5)$E-09	\\
$3s4p~^3P_{  2  }^o$ 	&	3.367E-09	&	3.375E-09	&	3.373E-09	&		&		&		\\
$3s4p~^1P_{  1  }^o$ 	&	1.972E-09	&	1.992E-09	&	1.926E-09	&		&		&		\\
$3p3d~^3F_{  2  }^o$ 	&	9.971E-07	&	1.160E-06	&	1.124E-06	&	0.023E-06	&		&		\\
$3p3d~^3F_{  3  }^o$ 	&	8.869E-07	&	1.004E-06	&	2.210E-06	&	0.025E-06	&		&		\\
$3p3d~^3F_{  4  }^o$ 	&	9.678E-07	&	1.272E-06	&	2.378E-06	&	0.026E-06	&		&		\\
$3s4d~^3D_{  3  }$   	&	2.871E-09	&	2.872E-09	&	2.818E-09	&		&	$(3.3 \pm 0.3)$E-09	&	 $(4.0 \pm 0.4)$E-09	\\
$3s4d~^3D_{  2  }$   	&	2.837E-09	&	2.846E-09	&	2.797E-09	&		&	$(3.3 \pm 0.3)$E-09	&	 $(4.0 \pm 0.4)$E-09	\\
$3s4d~^3D_{  1  }$   	&	2.814E-09	&	2.829E-09	&	2.783E-09	&		&	$(3.3 \pm 0.3)$E-09	&	 $(4.0 \pm 0.4)$E-09 	\\
$3s4d~^1D_{  2  }$   	&	1.263E-09	&	1.270E-09	&	1.294E-09	&		&	$(1.25 \pm 0.15)$E-09	&	 $(1.9 \pm 0.3)$E-09	\\
$3s4f~^1F_{  3  }^o$ 	&	6.169E-10	&	6.185E-10	&	5.979E-10	&		&		&		\\
$3p3d~^1D_{  2  }^o$ 	&	4.277E-10	&	4.273E-10	&	4.263E-10	&	3.800E-10	&		&		\\
$3s5s~^3S_{  1  }$   	&	7.438E-10	&	7.467E-10	&		&		&		&		\\
$3s5s~^1S_{  0  }$   	&	1.111E-09	&	1.114E-09	&		&		&		&		\\
$3s4f~^3F_{  2  }^o$ 	&	4.795E-10	&	4.788E-10	&		&		&	$(5.1 \pm 0.3)$E-10$^f$	&	$(12 \pm 1)$E-10	\\
$3s4f~^3F_{  3  }^o$ 	&	4.793E-10	&	4.786E-10	&		&		&	$(5.1 \pm 0.3)$E-10$^f$	&	$(12 \pm 1)$E-10	\\
$3s4f~^3F_{  4  }^o$ 	&	4.790E-10	&	4.780E-10	&		&		&	$(5.1 \pm 0.3)$E-10$^f$	&	$(12 \pm 1)$E-10	\\
$3s5p~^1P_{  1  }^o$ 	&	1.279E-09	&	1.288E-09	&		&		&		&		\\
$3s5p~^3P_{  2  }^o$ 	&	2.270E-09	&	2.276E-09	&		&		&		&		\\
$3s5p~^3P_{  1  }^o$ 	&	2.875E-09	&	2.889E-09	&		&		&		&		\\
$3s5p~^3P_{  0  }^o$ 	&	3.253E-09	&	3.273E-09	&		&		&		&		\\
$3p3d~^3P_{  2  }^o$ 	&	3.216E-10	&	3.225E-10	&		&		&		&		\\
$3p3d~^3P_{  1  }^o$ 	&	3.096E-10	&	3.110E-10	&		&		&		&		\\
$3p3d~^3P_{  0  }^o$ 	&	3.056E-10	&	3.076E-10	&		&		&		&		\\
$3p3d~^3D_{  1  }^o$ 	&	2.016E-10	&	2.025E-10	&		&		&		&	$(3.6 \pm 0.4)$E-10	\\
$3p3d~^3D_{  3  }^o$ 	&	2.020E-10	&	2.023E-10	&		&		&		&	$(3.6 \pm 0.4)$E-10	\\
$3p3d~^3D_{  2  }^o$ 	&	2.023E-10	&	2.028E-10	&		&		&		&	$(3.6 \pm 0.4)$E-10	\\
$3s5f~^1F_{  3~a}^o$  	    &	4.807E-10	&	4.824E-10	&		&		&		&	    $(10 \pm 5)$E-10	\\
$3p4s~^3P_{  0  }^o$ 	&	4.548E-10	&	4.538E-10	&		&		&		&		\\
$3p4s~^3P_{  1  }^o$ 	&	4.520E-10	&	4.511E-10	&		&		&		&		\\
$3p4s~^3P_{  2  }^o$ 	&	4.474E-10	&	4.468E-10	&		&		&		&		\\
$3s5d~^3D_{  3  }$   	&	8.682E-09	&	8.748E-09	&		&		&		&		\\
$3s5d~^3D_{  2  }$   	&	8.646E-09	&	8.735E-09	&		&		&		&		\\
$3s5d~^3D_{  1  }$   	&	8.617E-09	&	8.734E-09	&		&		&		&		\\
$3s5d~^1D_{  2  }$   	&	5.794E-09	&	5.833E-09	&		&		&		&		\\
$3p4s~^1P_{  1  }^o$ 	&	4.245E-10	&	4.246E-10	&		&		&		&		\\
$3s6s~^3S_{  1  }$   	&	1.285E-09	&	1.291E-09	&		&		&		&		\\
$3s5f~^3F_{  3  }^o$ 	&	9.537E-10	&	9.544E-10	&		&		&		&	$(14 \pm 2)$E-10	\\
$3s5f~^3F_{  2  }^o$ 	&	9.555E-10	&	9.566E-10	&		&		&		&	$(14 \pm 2)$E-10	\\
$3s5f~^3F_{  4  }^o$ 	&	9.511E-10	&	9.517E-10	&		&		&		&	$(14 \pm 2)$E-10	\\
$3s5g~^3G_{  3  }$   	&	2.775E-09	&	2.772E-09	&		&		&		&	$(4.3 \pm 0.5)$E-09	\\
$3s5g~^3G_{  4  }$   	&	2.776E-09	&	2.773E-09	&		&		&		&	$(4.3 \pm 0.5)$E-09	\\
$3s5g~^3G_{  5  }$   	&	2.775E-09	&	2.773E-09	&		&		&		&	$(4.3 \pm 0.5)$E-09	\\
$3s5g~^1G_{  4  }$   	&	2.779E-09	&	2.774E-09	&		&		&		&		\\
$3s6s~^1S_{  0  }$   	&	1.810E-09	&	1.815E-09	&		&		&		&		\\
$3s6p~^3P_{  0  }^o$ 	&	9.157E-09	&	9.335E-09	&		&		&		&		\\
$3s6p~^3P_{  1  }^o$ 	&	8.981E-09	&	9.117E-09	&		&		&		&		\\
$3s6p~^3P_{  2  }^o$ 	&	9.032E-09	&	9.105E-09	&		&		&		&		\\
$3p3d~^1P_{  1  }^o$ 	&	2.809E-10	&	2.843E-10	&		&		&		&		\\
$3s5f~^1F_{  3~b}^o$ 	&	3.995E-10	&	4.003E-10	&		&		&		&		\\
$3s6p~^1P_{  1  }^o$ 	&	1.129E-09	&	1.136E-09	&		&		&		&		\\
$3s6d~^3D_{  1  }$   	&	5.243E-09	&	5.354E-09	&		&		&		&		\\
$3s6d~^3D_{  2  }$   	&	5.548E-09	&	5.645E-09	&		&		&		&		\\
$3s6d~^3D_{  3  }$   	&	5.953E-09	&	6.035E-09	&		&		&		&		\\
$3s6d~^1D_{  2  }$   	&	1.081E-08	&	1.095E-08	&		&		&		&		\\
$3s7s~^3S_{  1  }$   	&	2.009E-09	&	2.021E-09	&		&		&		&		\\
$3s6f~^3F_{  4  }^o$ 	&	1.676E-09	&	1.679E-09	&		&		&		&		\\
$3s6f~^3F_{  3  }^o$ 	&	1.684E-09	&	1.687E-09	&		&		&		&		\\
$3s6f~^3F_{  2  }^o$ 	&	1.690E-09	&	1.695E-09	&		&		&		&		\\
$3s6g~^3G_{  3  }$   	&	4.764E-09	&	4.748E-09	&		&		&		&		\\
$3s6g~^3G_{  4  }$   	&	4.782E-09	&	4.763E-09	&		&		&		&		\\
$3s6g~^1G_{  4  }$   	&	4.802E-09	&	4.793E-09	&		&		&		&		\\
$3s6g~^3G_{  5  }$   	&	4.759E-09	&	4.754E-09	&		&		&		&		\\
$3s7s~^1S_{  0  }$   	&	2.849E-09	&	2.859E-09	&		&		&		&		\\
$3p4p~^1P_{  1  }$   	&	6.436E-10	&	6.446E-10	&		&		&		&		\\
$3s6f~^1F_{  3  }^o$ 	&	8.651E-10	&	8.659E-10	&		&		&		&		\\
$3p4p~^3D_{  1  }$   	&	7.732E-10	&	7.724E-10	&		&		&		&		\\
$3s7p~^3P_{  0  }^o$ 	&	1.524E-08	&	1.561E-08	&		&		&		&		\\
$3s7p~^3P_{  1  }^o$ 	&	1.521E-08	&	1.551E-08	&		&		&		&		\\
$3s7p~^3P_{  2  }^o$ 	&	1.518E-08	&	1.535E-08	&		&		&		&		\\
$3p4p~^3D_{  2  }$   	&	7.715E-10	&	7.714E-10	&		&		&		&		\\
$3p4p~^3D_{  3  }$   	&	7.665E-10	&	7.664E-10	&		&		&		&		\\
$3s7p~^1P_{  1  }^o$ 	&	7.755E-09	&	7.978E-09	&		&		&		&		\\
$3s7d~^1D_{  2  }$   	&	3.670E-09	&	3.710E-09	&		&		&		&		\\
$3p4p~^3P_{  0  }$   	&	7.308E-10	&	7.301E-10	&		&		&		&		\\
$3p4p~^3P_{  1  }$   	&	7.334E-10	&	7.326E-10	&		&		&		&		\\
$3p4p~^3P_{  2  }$   	&	7.504E-10	&	7.493E-10	&		&		&		&		\\
$3p4p~^3S_{  1  }$   	&	9.027E-10	&	9.046E-10	&		&		&		&		\\
$3s7d~^3D_{  1  }$   	&	6.279E-09	&	6.118E-09	&		&		&		&		\\
$3s7d~^3D_{  2  }$   	&	5.964E-09	&	5.850E-09	&		&		&		&		\\
$3s7d~^3D_{  3  }$   	&	5.550E-09	&	5.492E-09	&		&		&		&		\\
$3s7f~^3F_{  4  }^o$ 	&	2.689E-09	&	2.697E-09	&		&		&		&		\\
$3s7f~^3F_{  3  }^o$ 	&	2.708E-09	&	2.718E-09	&		&		&		&		\\
$3s7f~^3F_{  2  }^o$ 	&	2.722E-09	&	2.736E-09	&		&		&		&		\\
$3s7g~^3G_{  3  }$   	&	7.614E-09	&	7.543E-09	&		&		&		&		\\
$3s7g~^3G_{  4  }$   	&	7.632E-09	&	7.616E-09	&		&		&		&		\\
$3s7g~^1G_{  4  }$   	&	7.804E-09	&	7.727E-09	&		&		&		&		\\
$3s7g~^3G_{  5  }$   	&	7.610E-09	&	7.609E-09	&		&		&		&		\\
$3s7f~^1F_{  3  }^o$ 	&	1.713E-09	&	1.723E-09	&		&		&		&		\\
\hline
\end{longtable}
\tablebib{
$^{(a)}$Present calculations; $^{(b)}$\citet{FroeseFischer06}; $^{(c)}$\citet{Safronova00}; $^{(d)}$\citet{Bashkin80}; $^{(e)}$\citet{Berry71}; $^{(f)}$\citet{Livingston76}.
}

%%%%%%%%%%%%%%%%%%%%%%%%%%%%%%%%%%%%%%%%%%%%%%%%%%%%%%%%%%%%

%%%%%%%%%%%% 	Energy Tables  Si IV 	%%%%%%%%%%%%%%%%%%%%%%%%%%%%%%%%%%%%

%%%%%%%%%%%%%%%%%%%%%%%%%%%%%%%%%%%%%%%%%%%%%%%%%%%%%%%%%%%%

\centering
\begin{longtable}{lllllllll} 
\caption{ \label{SiIV_energy} Computed excitation energies in cm $^{-1}$ for the 45 lowest states in \ion{Si}{iv}, as a function of the increasing active set of orbitals, accounting for CV correlation, where $n$ indicates the maximum principle quantum number of the orbitals
included in the active set, as well as CC correlation. Observed and computed excitation energies for the \ion{Si}{iv} from the NIST-database ($E_{\text{obs}}^a$) and the other theoretical study ($E_{\text{theor}}^{b}$) are also given. In the last two columns, the difference $\Delta E $ between observed and computed energies is compared for the present and previous studies.}\\ \hline\hline 
&& CV &&&& \\
\cline{2-4}
Level  	&	$n=10$	        &	$n=11$	        &	$n=12$	        &	CC	        &	$E_{\text{obs}}$$^a$	& $E_{\text{theor}}^b$       &	$\Delta E_{\text{CC}}$ 	&	$\Delta E_{\text{theor}}^b$ \\
\hline
$3s~^2S_{  1/2  }$  ~  	&	0	&	0	&	0	&	0	&	0	&	0	&	0	&	0	\\
$3p~^2P_{  1/2  }^o$~  	&	71407	&	71353	&	71351	&	71211	&	71288	&	71159	&	77	&	128	\\
$3p~^2P_{  3/2  }^o$~  	&	71870	&	71813	&	71812	&	71667	&	71749	&	71667	&	82	&	81	\\
$3d~^2D_{  5/2  }$  ~  	&	160585	&	161039	&	161209	&	160208	&	160374	&	160287	&	166	&	87	\\
$3d~^2D_{  3/2  }$  ~  	&	160587	&	161041	&	161210	&	160210	&	160376	&	160281	&	166	&	94	\\
$4s~^2S_{  1/2  }$  ~  	&	194179	&	194571	&	194709	&	193778	&	193979	&	193787	&	201	&	192	\\
$4p~^2P_{  1/2  }^o$~  	&	218498	&	218859	&	218986	&	218048	&	218267	&	218051	&	219	&	216	\\
$4p~^2P_{  3/2  }^o$~  	&	218661	&	219020	&	219147	&	218208	&	218429	&	218234	&	221	&	194	\\
$4d~^2D_{  5/2  }$  ~  	&	250200	&	250743	&	250933	&	249767	&	250008	&	249827	&	241	&	181	\\
$4d~^2D_{  3/2  }$  ~  	&	250200	&	250743	&	250933	&	249768	&	250008	&	249821	&	240	&	187	\\
$4f~^2F_{  5/2  }^o$~  	&	254313	&	254927	&	255134	&	253847	&	254127	&	253876	&	280	&	251	\\
$4f~^2F_{  7/2  }^o$~  	&	254315	&	254929	&	255136	&	253848	&	254129	&	253880	&	281	&	249	\\
$5s~^2S_{  1/2  }$  ~  	&	265617	&	266135	&	266312	&	265166	&	265418	&	265216	&	252	&	202	\\
$5p~^2P_{  1/2  }^o$~  	&	276718	&	277218	&	277387	&	276244	&	276504	&	276281	&	260	&	223	\\
$5p~^2P_{  3/2  }^o$~  	&	276794	&	277293	&	277462	&	276319	&	276579	&	276365	&	260	&	214	\\
$5d~^2D_{  5/2  }$  ~  	&	291680	&	292265	&	292464	&	291232	&	291498	&	291292	&	266	&	206	\\
$5d~^2D_{  3/2  }$  ~  	&	291680	&	292265	&	292464	&	291232	&	291498	&	291290	&	266	&	208	\\
$5f~^2F_{  5/2  }^o$~  	&	293898	&	294519	&	294727	&	293431	&	293719	&	293475	&	288	&	244	\\
$5f~^2F_{  7/2  }^o$~  	&	293899	&	294520	&	294728	&	293432	&	293719	&	293477	&	287	&	242	\\
$5g~^2G_{  7/2  }$  ~  	&	294007	&	294638	&	294847	&	293549	&	293838	&		&	289	&		\\
$5g~^2G_{  9/2  }$  ~  	&	294008	&	294638	&	294848	&	293550	&	293838	&		&	288	&		\\
$6s~^2S_{  1/2  }$  ~  	&	299865	&	300436	&	300628	&	299406	&	299677	&	299451	&	271	&	226	\\
$6p~^2P_{  1/2  }^o$~  	&	305839	&	306398	&	306585	&	305365	&	305641	&	305405	&	276	&	236	\\
$6p~^2P_{  3/2  }^o$~  	&	305881	&	306439	&	306626	&	305406	&	305682	&	305450	&	276	&	232	\\
$6d~^2D_{  3/2  }$  ~  	&	314092	&	314698	&	314901	&	313638	&	313915	&	313688	&	277	&	227	\\
$6d~^2D_{  5/2  }$  ~  	&	314092	&	314698	&	314901	&	313638	&	313915	&	313690	&	277	&	225	\\
$6f~^2F_{  5/2  }^o$~  	&	315404	&	316031	&	316239	&	314939	&	315230	&	314983	&	291	&	248	\\
$6f~^2F_{  7/2  }^o$~  	&	315404	&	316031	&	316240	&	314940	&	315230	&	314983	&	290	&	247	\\
$6g~^2G_{  7/2  }$  ~  	&	315475	&	316106	&	316315	&	315016	&	315305	&		&	289	&		\\
$6g~^2G_{  9/2  }$  ~  	&	315475	&	316106	&	316316	&	315016	&	315305	&		&	289	&		\\
$6h~^2H_{  9/2  }^o$~  	&	315482	&	316118	&	316327	&	315023	&	315317	&		&	294	&		\\
$6h~^2H_{ 11/2  }^o$~  	&	315482	&	316118	&	316328	&	315023	&	315317	&		&	294	&		\\
$7s~^2S_{  1/2  }$  ~  	&	318924	&	319520	&	319719	&	318463	&	318743	&	318505	&	280	&	238	\\
$7p~^2P_{  1/2  }^o$~  	&	322500	&	323089	&	323285	&	322029	&	322313	&	322069	&	284	&	244	\\
$7p~^2P_{  3/2  }^o$~  	&	322525	&	323114	&	323309	&	322054	&	322338	&	322097	&	284	&	241	\\
$7d~^2D_{  3/2  }$  ~  	&	327536	&	328153	&	328358	&	327080	&	327362	&	327125	&	282	&	237	\\
$7d~^2D_{  5/2  }$  ~  	&	327536	&	328153	&	328358	&	327080	&	327362	&	327125	&	282	&	237	\\
$7f~^2F_{  5/2  }^o$~  	&	328370	&	329001	&	329210	&	327908	&	328200	&	327950	&	292	&	251	\\
$7f~^2F_{  7/2  }^o$~  	&	328371	&	329001	&	329210	&	327908	&	328200	&	327950	&	292	&	250	\\
$7g~^2G_{  7/2  }$  ~  	&	328418	&	329050	&	329260	&	327959	&	328250	&		&	291	&		\\
$7g~^2G_{  9/2  }$  ~  	&	328419	&	329051	&	329260	&	327959	&	328250	&		&	291	&		\\
$7h~^2H_{  9/2  }^o$~  	&	328423	&	329059	&	329269	&	327963	&	328257	&		&	294	&		\\
$7h~^2H_{ 11/2  }^o$~  	&	328423	&	329059	&	329269	&	327964	&	328257	&		&	293	&		\\
$7i~^2I_{ 11/2  }$  ~  	&	328424	&	329060	&	329270	&	327965	&	328261	&		&	296	&		\\
$7i~^2I_{ 13/2  }$  ~  	&	328425	&	329061	&	329270	&	327965	&	328261	&		&	296	&		\\
\hline\\
\end{longtable}
\tablebib{
$^{(a)}$\citet{NIST_ASD}; $^{(b)}$\citet{FroeseFischer06}.
}

%%%%%%%%%%%%%%%%%%%%%%%%%%%%%%%%%%%%%%%%%%%%%%%%%%%%%%%%%%%%

%%%%%%%%%%%% Lifetime Tables  Si IV	%%%%%%%%%%%%%%%%%%%%%%%%%%%%%%%%%%%%

%%%%%%%%%%%%%%%%%%%%%%%%%%%%%%%%%%%%%%%%%%%%%%%%%%%%%%%%%%%%

%\begin{longtab}
\centering
\begin{longtable}{lllllll} 
\caption{\label{SiIV_lifetime_comp} Results for \ion{Si}{IV}: Comparison between computed lifetimes, in length ($\tau_l$) and velocity ($\tau_v$) gauge, from our calculations. These are compared to the predicted lifetimes from other calculations$^{b,c}$ ($\tau_{\text{theor}}$), as well as experimental results $\tau_{\text{obs}}^{d,e}$, with stated uncertainties. All values are given in seconds.  }\\ \hline\hline
&&RCI&&&&\\
\cline{2-3}
Level & $\tau_l$ & $\tau_v$ & $\tau_{\text{theor}}$ $^b$& $\tau_{\text{theor}}$$^c$ & $\tau_{\text{obs}}$$^d$ &  $\tau_{\text{obs}}$$^e$\\ \hline

$3p~^2P_{ 1/2  }^o$ 	&	1.162E-09	&	1.159E-09	&	1.167E-09	&	1.102E-09	&		&	$(1.2 \pm 0.4)$E-09	\\
$3p~^2P_{ 3/2  }^o$ 	&	1.139E-09	&	1.138E-09	&	1.141E-09	&	1.081E-09	&		&	$(1.2 \pm 0.4)$E-09	\\
$3d~^2D_{ 3/2  }$   	&	3.923E-10	&	3.906E-10	&	3.916E-10	&	3.790E-10	&		&	$(4.6 \pm 0.5)$E-10	\\
$3d~^2D_{ 5/2  }$   	&	3.964E-10	&	3.949E-10	&	3.962E-10	&	3.840E-10	&		&	$(4.6 \pm 0.5)$E-10	\\
$4s~^2S_{ 1/2  }$   	&	2.819E-10	&	2.827E-10	&	2.821E-10	&	2.930E-10	&		&		\\
$4p~^2P_{ 1/2  }^o$ 	&	9.156E-10	&	9.150E-10	&	9.111E-10	&	9.970E-10	&		&	$(7.5 \pm 0.3)$E-10	\\
$4p~^2P_{ 3/2  }^o$ 	&	9.232E-10	&	9.221E-10	&	9.199E-10	&	9.900E-10	&		&	$(7.5 \pm 0.3)$E-10	\\
$4d~^2D_{ 3/2  }$   	&	1.808E-09	&	1.812E-09	&	1.806E-09	&	1.824E-09	&	$(2.0 \pm 0.2)$E-09	&	$(2.1 \pm 0.2)$E-09	\\
$4d~^2D_{ 5/2  }$   	&	1.811E-09	&	1.816E-09	&	1.810E-09	&	1.846E-09	&	$(2.0 \pm 0.2)$E-09	&	$(2.1 \pm 0.2)$E-09	\\
$4f~^2F_{ 5/2  }^o$ 	&	2.638E-10	&	2.638E-10	&	2.639E-10	&	2.620E-10	&		&	$(4.8 \pm 0.4)$E-10	\\
$4f~^2F_{ 7/2  }^o$ 	&	2.638E-10	&	2.638E-10	&	2.639E-10	&	2.620E-10	&		&	$(4.8 \pm 0.4)$E-10	\\
$5s~^2S_{ 1/2  }$   	&	4.348E-10	&	4.356E-10	&	4.322E-10	&	4.460E-10	&		&		\\
$5p~^2P_{ 1/2  }^o$ 	&	1.288E-09	&	1.283E-09	&	1.269E-09	&	1.382E-09	&	$(2.1 \pm 0.3)$E-09	&	$(2.2 \pm 0.2)$E-09	\\
$5p~^2P_{ 3/2  }^o$ 	&	1.299E-09	&	1.293E-09	&	1.283E-09	&	1.375E-09	&	$(2.1 \pm 0.3)$E-09	&	$(2.2 \pm 0.2)$E-09	\\
$5d~^2D_{ 3/2  }$   	&	3.090E-09	&	3.100E-09	&	3.058E-09	&	3.231E-09	&		&		\\
$5d~^2D_{ 5/2  }$   	&	3.068E-09	&	3.078E-09	&	3.036E-09	&	3.256E-09	&		&		\\
$5f~^2F_{ 5/2  }^o$ 	&	4.940E-10	&	4.934E-10	&	4.926E-10	&		&		&		\\
$5f~^2F_{ 7/2  }^o$ 	&	4.940E-10	&	4.935E-10	&	4.927E-10	&	4.820E-10	&		&		\\
$5g~^2G_{ 7/2  }$   	&	9.147E-10	&	9.154E-10	&		&	9.240E-10	&	$(12 \pm 2)$E-10	&	$(26 \pm 3)$E-10	\\
$5g~^2G_{ 9/2  }$   	&	9.147E-10	&	9.154E-10	&		&	9.240E-10	&	$(12 \pm 2)$E-10	&	$(26 \pm 3)$E-10	\\
$6s~^2S_{ 1/2  }$   	&	7.062E-10	&	7.058E-10	&	6.952E-10	&	7.160E-10	&		&		\\
$6p~^2P_{ 1/2  }^o$ 	&	1.957E-09	&	1.937E-09	&	1.901E-09	&	2.063E-09	&		&		\\
$6p~^2P_{ 3/2  }^o$ 	&	1.975E-09	&	1.952E-09	&	1.921E-09	&	2.056E-09	&		&		\\
$6d~^2D_{ 3/2  }$   	&	4.529E-09	&	4.488E-09	&	4.376E-09	&	4.685E-09	&		&		\\
$6d~^2D_{ 5/2  }$   	&	4.482E-09	&	4.445E-09	&	4.332E-09	&	4.712E-09	&		&		\\
$6f~^2F_{ 5/2  }^o$ 	&	8.336E-10	&	8.289E-10	&	8.274E-10	&		&		&		\\
$6f~^2F_{ 7/2  }^o$ 	&	8.336E-10	&	8.290E-10	&	8.274E-10	&	8.070E-10	&		&		\\
$6g~^2G_{ 7/2  }$   	&	1.568E-09	&	1.569E-09	&		&		&		&		\\
$6g~^2G_{ 9/2  }$   	&	1.568E-09	&	1.569E-09	&		&	1.569E-09	&		&		\\
$6h~^2H_{ 9/2  }^o$ 	&	2.373E-09	&	2.374E-09	&		&		&		&	$(5.2 \pm 0.5)$E-09	\\
$6h~^2H_{11/2  }^o$ 	&	2.373E-09	&	2.374E-09	&		&	2.376E-09	&		&	$(5.2 \pm 0.5)$E-09	\\
$7s~^2S_{ 1/2  }$   	&	1.116E-09	&	1.110E-09	&	1.077E-09	&	1.110E-09	&		&		\\
$7p~^2P_{ 1/2  }^o$ 	&	3.010E-09	&	2.911E-09	&	2.808E-09	&	3.040E-09	&		&		\\
$7p~^2P_{ 3/2  }^o$ 	&	3.035E-09	&	2.934E-09	&	2.837E-09	&	3.034E-09	&		&		\\
$7d~^2D_{ 3/2  }$   	&	6.832E-09	&	6.366E-09	&	6.073E-09	&	6.542E-09	&		&		\\
$7d~^2D_{ 5/2  }$   	&	6.732E-09	&	6.300E-09	&	6.007E-09	&	6.553E-09	&		&		\\
$7f~^2F_{ 5/2  }^o$ 	&	1.325E-09	&	1.294E-09	&	1.289E-09	&		&		&		\\
$7f~^2F_{ 7/2  }^o$ 	&	1.325E-09	&	1.294E-09	&	1.289E-09	&	1.261E-09	&		&		\\
$7g~^2G_{ 7/2  }$   	&	2.473E-09	&	2.469E-09	&		&		&		&		\\
$7g~^2G_{ 9/2  }$   	&	2.473E-09	&	2.470E-09	&		&	2.454E-09	&		&		\\
$7h~^2H_{ 9/2  }^o$ 	&	3.751E-09	&	3.752E-09	&		&		&		&		\\
$7h~^2H_{11/2  }^o$ 	&	3.751E-09	&	3.753E-09	&		&	3.750E-09	&		&		\\
$7i~^2I_{11/2  }$   	&	5.270E-09	&	5.272E-09	&		&		&		&		\\
$7i~^2I_{13/2  }$   	&	5.270E-09	&	5.272E-09	&		&	5.267E-09	&		&		\\
\hline\\
\end{longtable}
\tablefoot{Both the length  and velocity forms are displayed for the present RCI calculations. The available lifetimes from experimental measurements and previous calculations are also given.  }
\tablebib{$^{(a)}$Present calculations; $^{(b)}$\citet{FroeseFischer06}; $^{(c)}$\citet{SIEMS2001}; $^{(d)}$\citet{Bashkin80}; $^{(e)}$\citet{Berry71}.}
%\end{longtab}

\end{appendix}

\end{document}